\documentclass[12pt]{article}
\usepackage{epsfig, amsmath, amssymb}
\usepackage{graphicx,psfrag}
\usepackage{xcolor}
\newcommand{\nn}{\nonumber}
\newcommand{\be}{\begin{equation}}
\newcommand{\ee}{\end{equation}}
\newcommand{\ben}{\begin{equation}}
\newcommand{\een}{\end{equation}}
\newcommand{\bea}{\begin{eqnarray}}
\newcommand{\eea}{\end{eqnarray}}
\newcommand{\bA}{\begin{array}}
\newcommand{\eA}{\end{array}}
\newcommand{\bc}{\begin{center}}
\newcommand{\ec}{\end{center}}
\newcommand{\al}{\alpha}

\newcommand{\ra}{\rightarrow}

\newcommand{\ie}{{\it i.e.}}
\newcommand{\eg}{{\it e.g.}}

\newcommand{\ran}{\rangle}

\begin{document}


\begin{titlepage}

%
  
\bc

\hfill 
\\         [22mm]

{\Huge Schwarzschild de Sitter and extremal surfaces}
\vspace{16mm}

{\large Karan Fernandes$^{1,2}$,\ \ Kedar S.~Kolekar$^{2,3}$,\ \
  K.~Narayan$^2$,\ \ Sourav Roy$^{2,4}$} \\
\vspace{3mm}
{\small \it 1. Harish-Chandra Research Institute\\}
{\small \it Chhatnag Road, Jhusi, Allahabad 211019, India.\\ [3mm]}

{\small \it 2. Chennai Mathematical Institute, \\}
{\small \it H1 SIPCOT IT Park, Siruseri 603103, India.\\ [3mm]}

{\small \it 3. Department of Physics, Indian Institute of Technology Kanpur,\\}
{\small \it Kanpur 208016, India.\\ [3mm]}

{\small \it 4. Department of Physics, Syracuse University, \\}
{\small \it Syracuse, NY, USA.\\ }

\ec
\vspace{25mm}

\begin{abstract}
  We study extremal surfaces in the Schwarzschild de Sitter spacetime
  with real mass parameter. We find codim-2 timelike extremal surfaces
  stretching between the future and past boundaries that pass through
  the vicinity of the cosmological horizon in a certain limit. These
  are analogous to the surfaces in arXiv:1711.01107 [hep-th]. We also
  find spacelike surfaces that never reach the future/past boundaries
  but stretch indefinitely through the extended Penrose diagram,
  passing through the vicinity of the cosmological and Schwarzschild
  horizons in a certain limit. Further, these exhibit interesting
  structure for de Sitter space (zero mass) as well as in the
  extremal, or Nariai, limit.
\end{abstract}


\end{titlepage}

{\tiny 
\begin{tableofcontents}
\end{tableofcontents}
}


\section{Introduction and summary}

Over the last several years, holographic entanglement entropy
\cite{Ryu:2006bv,Ryu:2006ef,HRT,Rangamani:2016dms} has been under
substantial investigation, both from the point of view of gaining new
insights on strongly coupled field theories as well as on spacetime
geometry intertwining with entanglement via gauge/gravity duality
\cite{Maldacena:1997re,Gubser:1998bc,Witten:1998qj,Aharony:1999ti}.
The RT/HRT proposals involve extremal surfaces whose area encodes
entanglement entropy in the dual field theory. In $AdS$, surfaces
anchored at one end of a subsystem dip into the bulk radial direction
upto the ``deepest'' location which is the ``turning point'', and then
begin to return to the boundary.

It is a fascinating question to extend these explorations to de Sitter
space (see \eg\ \cite{Spradlin:2001pw} for a review), which is known
to possess entropy \cite{Gibbons:1977mu}, given by the area of the
cosmological horizon.  One might imagine this dovetails with attempts
to understand de Sitter entropy via gauge/gravity duality for de
Sitter space, or $dS/CFT$
\cite{Strominger:2001pn,Witten:2001kn,Maldacena:2002vr}, which
conjecture $dS$ to be dual to a hypothetical Euclidean non-unitary
Conformal Field Theory that lives on ${\cal I}^+$, with the dictionary
$\Psi_{dS}=Z_{CFT}$\ \cite{Maldacena:2002vr}.\ \ $\Psi_{dS}$ is the
late-time Hartle-Hawking Wavefunction of the Universe with appropriate
boundary conditions and $Z_{CFT}$ the dual CFT partition function.
Dual energy-momentum tensor correlation functions reveal a negative
central charge $-{R_{dS}^2\over G_4}$ for $dS_4$
suggesting a ghost-like $CFT_3$ dual: this is exemplified in the higher
spin $dS_4$ duality involving a 3-dim CFT of anticommuting (ghost)
scalars \cite{Anninos:2011ui}. Bulk expectation values
\cite{Maldacena:2002vr} are obtained as\
$\langle \varphi_k \varphi_{k'}\rangle  \sim \int D\varphi\ \varphi_k
\varphi_{k'} |\Psi_{dS}|^2$, weighting with the bulk probability
$|\Psi_{dS}|^2=\Psi_{dS}^*\Psi_{dS}$: the presence of $\Psi_{dS}$ and
$\Psi_{dS}^*$ suggests that bulk de Sitter physics involves two copies
of the dual CFT, possibly on the future and past boundaries.

In this context, it is interesting to ask if de Sitter entropy is some
sort of generalized entanglement entropy via $dS/CFT$
\cite{Narayan:2017xca}: a class of investigations in this regard is
summarized in \cite{Narayan:2019pjl}.  Various explorations of
extremal surfaces in de Sitter space were carried out in
\cite{Narayan:2015vda} in the Poincare slicing and in
\cite{Narayan:2017xca} in the static coordinatization, looking for
extremal surfaces anchored at $I^+$, since the natural boundary for
$dS$ is future/past timelike infinity $I^\pm$. Since the dual CFT is
Euclidean and there is no ``intrinsic'' boundary time, operationally
we take some spatial isometry direction as boundary Euclidean time, as
a crutch: de Sitter isometries imply that no particular slice is
sacrosanct. We then define a subsystem on this slice, which leads to
codim-2 bulk extremal surfaces stretching in the time direction. In
general it then appears that real surfaces do not exhibit any turning
point where a surface starting at $I^+$ begins to return to $I^+$.
(There are complex extremal surfaces with turning points that amount
to analytic continuation from the Ryu-Takayanagi $AdS$ expressions
\cite{Narayan:2015vda}, giving negative area in
$dS_4$ consistent with the negative central charge; see also
\cite{Sato:2015tta,Miyaji:2015yva}. However their interpretation is
not entirely clear.)\ It is thus interesting to ask if surfaces
beginning at $I^+$ stretch all the way to $I^-$: in
\cite{Narayan:2017xca}, such codim-2 real timelike surfaces were in
fact found.
These exhibit an area law divergence ${\pi l^2\over G_4}
{1\over\epsilon}$, where $\epsilon={\epsilon_c\over l}$ is the
dimensionless ultraviolet cutoff (with $l$ the de Sitter scale) and
the coefficient scales as de Sitter entropy.  In more detail, the
static patch coordinatization can be recast as
\be\label{dSmetric-tau}
ds^2 = {l^2\over\tau^2} \Big(-{d\tau^2\over 1-\tau^2} + (1-\tau^2) dw^2
+ d\Omega_{d-1}^2\Big)\ ,
\ee
with the future/past universes $F/P$ parametrized by $0\leq\tau\leq 1$
with horizons at $\tau=1$, while the Northern/Southern diamonds $N/S$
have $1<\tau\leq\infty$. The boundaries at $\tau=0$ are now of the
form $R_w\times S^{d-1}$, resembling the Poincare slicing locally.
Setting up the extremization for codim-2 surfaces on boundary
Euclidean time slices can be carried out: on $S^{d-1}$ equatorial
planes for instance we obtain\ ${\dot w}^2 = {B^2\tau^{2d-2}\over
  1-\tau^2 + B^2\tau^{2d-2}}$\ with real turning points in the 
Northern/Southern diamonds ($N/S$). These connected surfaces are analogous to
rotated versions of surfaces found by Hartman, Maldacena
\cite{Hartman:2013qma} in the AdS black hole, argued to be dual to a
thermofield double state \cite{Maldacena:2001kr}. The existence of the
connected surfaces stretching between $I^\pm$, in light of the fact
that the bulk de Sitter space has entropy, suggests the speculation
\cite{Narayan:2017xca} that $dS_4$ is approximately dual to an
entangled thermofield-double type state\ $|\psi\rangle = \sum
\psi^{i_n^F,i_n^P} |i_n^F\ran |i_n^P\ran$ in two copies $CFT_F\times
CFT_P$ of the ghost-CFT, at $I^+$ and $I^-$, the generalized
entanglement entropy of the latter scaling as de Sitter entropy. (See
also \cite{Arias:2019pzy} in this regard.) This is also in part
motivated by parallel investigations of certain generalizations of
entanglement to ghost-like theories, in particular ghost-spins
\cite{Narayan:2016xwq}, where ``correlated'' states of this kind,
entangling identical states $i_n^F$ and $i_n^P$ in two copies of
ghost-spin ensembles, were found to uniformly have positive norm,
reduced density matrix and entanglement.

In light of the bulk extremal surface studies above, it is of interest
to explore the Schwarzschild de Sitter spacetime. This represents a
Schwarzschild black hole in de Sitter space \cite{Gibbons:1977mu}, and
thus exhibits a cosmological horizon as well as a Schwarzschild
horizon, for certain ranges $0< {m\over l} \leq {1\over\sqrt{27}}$\ of
the dimensionless mass parameter ${m\over l}$ where $l$ is the de
Sitter scale. An interesting limit here arises for ${m\over l} =
{1\over\sqrt{27}}$\,: in this extremal or Nariai limit \cite{Nariai},
the region between the horizons becomes $dS_2\times
S^2$. Schwarzschild de Sitter spacetimes also arise as the dominant
contributions to the late-time Hartle-Hawking wavefunction for
asymptotically $S^1\times S^2$ geometries in certain limits
\cite{Anninos:2012ft}, but with imaginary mass parameter; see also
\cite{Maldacena:2019cbz} for more on the nearly $dS_2$ limit, the
wavefunction and the no-boundary proposal. In what follows, we will
mostly be interested in the real mass case: unfortunately this does
not appear to have bearing on $dS_4/CFT_3$ so we are mainly studying
this as a gravitational question alone. It is then of interest to look
for codim-2 real timelike surfaces stretching from $I^+$ to
$I^-$. Along the lines of \cite{Narayan:2017xca}, we do find codim-2
real timelike extremal surfaces passing through the vicinity of the
cosmological horizon.  These are described by\ ${\dot w}^2 =
{B^2\tau^{2d-2}\over f(\tau) + B^2\tau^{2d-2}}$\,, where
$f(\tau)=1-\tau^2+{2m\over l}\tau^d$ is the metric factor and $B^2>0$
is a conserved quantity that is a parameter encoding the width
boundary conditions at $I^\pm$ as in the de Sitter case. In the limit
where the subregion at $I^\pm$ becomes the whole space, we find a
limiting surface as in the de Sitter case (described in detail in
\cite{Narayan:2020nsc}). 
As $m\ra 0$, this analysis coincides with the de Sitter case. Along
similar lines, one might expect to find similar surfaces passing
through the vicinity of the Schwarzschild horizon: we do not find any
real surfaces of this kind though, as we explain in detail (sec.~2).

However we do find that surfaces of the above kind with $B^2<0$ in
fact exhibit interesting behaviour (sec.~3). These end up being
spacelike surfaces passing through the vicinity of the Schwarzschild
and cosmological horizons in a certain limit. They extend indefinitely
through the extended Penrose diagram, and never reach any
$I^\pm$. These spacelike surfaces do admit interesting limits in the
extremal or Nariai limit of Schwarzschild de Sitter, as we
discuss. The de Sitter limit of these spacelike surfaces interestingly
has area equal to de Sitter entropy.  Finally we show that these
spacelike surfaces can also be obtained as certain analytic
continuations of certain extremal surfaces in global $AdS$. Some
appendices discuss some technical aspects of the tortoise coordinate
and Penrose diagrams, an analysis of the 3-dim Schwarzschild de Sitter
spacetime which is a bit special, and some technical aspects of the
area integrals that arise in the paper.

\section{Schwarzschild de Sitter and extremal surfaces}

The Schwarzschild de Sitter spacetime in $d+1$-dimensions is given
by the metric
\be\label{SdSst}
ds^2= -f(r)dt^2+\frac{dr^2}{f(r)}+r^2d\Omega_{d-1}^2\ ,  \qquad  
f(r)=1-\frac{2m}{l}\left(\frac{l}{r}\right)^{d-2}-\frac{r^2}{l^2}\ ,
\ee
and describes a Schwarzschild black hole in de Sitter space
\cite{Gibbons:1977mu} with an ``outer'' cosmological horizon as well
as an ``inner'' Schwarzschild horizon. The surface gravity at both
horizons is distinct generically and Euclidean continuations can be
defined removing a conical singularity at either horizon but not both
simultaneously \cite{Ginsparg:1982rs} (see also
\cite{Bousso:1995cc,Bousso:1996au}). The periodicities of Euclidean
time coincide in an extremal, or Nariai, limit \cite{Nariai}, which is
degenerate: away from this precise value, the periodicities cannot
match. The spacetime develops a nearly $dS_2$ throat in a near
extremal limit \cite{Ginsparg:1982rs}. Schwarzschild de Sitter
spacetimes with large and purely imaginary mass\footnote{A related
  spacetime was found in \cite{Das:2013mfa}, representing a $dS_4$
  black brane: for imaginary energy density parameter, this leads via
  $dS/CFT$ to real energy-momentum density in the dual CFT. For the
  parameter real, the spacetime has a Penrose diagram that resembles
  the interior of the Reissner-Nordstrom black hole, exhibiting
  timelike singularities cloaked by Cauchy horizons which give rise to
  a blueshift instability. These do not admit a Nariai limit.} give
the dominant contribution to the finite part of the late time
Hartle-Hawking wavefunction of the universe for asymptotically
$S^1\times S^2$ geometries satisfying the no-boundary proposal in the
limit where the $S^1$ size is small \cite{Anninos:2012ft}. More on the
nearly $dS_2$ limit and the wavefunction of the universe appears in
\cite{Maldacena:2019cbz}.

As for the de Sitter case in \cite{Narayan:2017xca}, we find it
useful to recast this metric in terms of the coordinates
$\tau=\frac{l}{r}, w=\frac{t}{l}$\,: this gives
\be\label{SdSst2}
ds^2 = {l^2\over\tau^2} \left(-{d\tau^2\over f(\tau)} + f(\tau) dw^2
+ d\Omega_{d-1}^2\right) , \qquad  
f(\tau) =  1-\tau^2+{2m\over l}\tau^d\ .
\ee
In the $\tau$-coordinate, the future boundary $I^+$ is at $\tau=0$ and
singularities arise at $\tau\ra\infty$.  Now near $I^+$, the metric
locally resembles the Poincare patch. The horizons are given by the
zeros of $f(\tau)$. The de Sitter limit with $m=0$ becomes
(\ref{dSmetric-tau}): there is a cosmological horizon at $\tau=1$.
For $d=3$ \ie\ $SdS_4$ with nonzero real mass $m$, $f(\tau)$ is a
cubic function with multiple zeros: two physically interesting roots
arise for $0\leq {m\over l}\leq {1\over\sqrt{27}}$ representing a
cosmological and a Schwarzschild horizon \cite{Gibbons:1977mu}, as we
describe below. The maximally extended Penrose diagram shows an
infinitely repeating pattern of ``unit-cells'', with cosmological
horizons bounding future/past universes $F/P$, Schwarzschild horizons
bounding interior regions $I_F, I_P$ and an intermediate diamond
region $D$.

We will mostly focus on the 4-dim Schwarzschild de Sitter case in what
follows\ (an appendix studies $SdS_3$ in detail).\ For $SdS_4$, we have
\bea\label{SdS4-fma1a2}
&& SdS_4:\qquad
f(\tau) = 1-\tau^2+{2m\over l}\tau^3
= (1-a_1\tau)(1-a_2\tau)(1+(a_1+a_2)\tau)\ ,
\nonumber\\
&&\qquad\qquad\qquad a_1^2+a_2^2+a_1a_2=1\ ,\qquad a_1a_2(a_1+a_2)={2m\over l}\ .
\eea
Thus the roots $a_1,a_2$ are constrained as above, and taking the
positive root for $a_2$ gives $a_2={1\over 2} (\sqrt{4-3a_1^2}-a_1)$\,.
Upto an overall $\tau^2$ factor, $f(\tau)$ is the same as $f(r)$ in
(\ref{SdSst}) so the zeroes of $f(\tau)$ give the locations of the
horizons.  Thus in the above, we have\ $\tau_c={1\over a_1}$ and
$\tau_s={1\over a_2}$ as the two physical values, corresponding to the
cosmological (de Sitter) and Schwarzschild horizons. (The third zero
does not correspond to a physical horizon.)  The case with $m=0$, or
$a_1=1, a_2=0$, is pure de Sitter space.  This structure of horizons
is valid for ${m\over l}<{1\over 3\sqrt{3}}$\,, beyond which there are
no horizons.  The limit ${m\over l}={1\over 3\sqrt{3}}$ corresponds to
the cosmological and Schwarzschild horizon values coinciding: here we
have $a_1=a_2=a_0={1\over\sqrt{3}}$\ from (\ref{SdS4-fma1a2}). This
special value leads to the extremal, or Nariai, limit where the near
horizon region (between the horizons) becomes $dS_2\times S^2$.
Overall the range of physically interesting $a_1, a_2$ satisfies\ $0 <
a_2 < a_0 < a_1$ for generic values, and\ ${1\over a_1} < {1\over
  a_2}$\ implies that the cosmological horizon is ``outside'' the
Schwarzschild one.

\begin{figure}[h] 
\includegraphics[width=16pc]{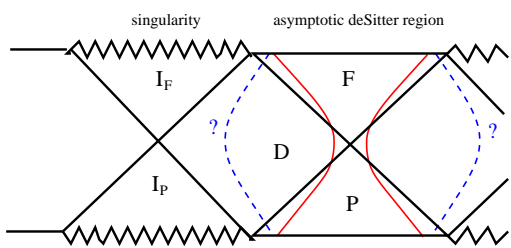} \hspace{1pc}
\begin{minipage}[b]{21pc}
  \caption{{\label{SdSsurf0}\footnotesize {\ \ Timelike extremal surfaces
        $w(\tau)$ in SdS stretching between $I^\pm$, shown as the red
        curves passing through the vicinity of the cosmological
        horizon. The dashed blue curves represent hypothetical
        timelike extremal surfaces of similar nature, but passing
        near the Schwarzschild horizon in some limit.
}}}
\end{minipage}
\end{figure}
We want to first consider codim-2 timelike extremal surfaces that
stretch from $I^+$ to $I^-$ within a ``unit-cell'', as shown in
Figure~\ref{SdSsurf0}. These lie in some equatorial plane of the $S^2$
and thus wrap an $S^1$ within $S^2$, extend from some fixed boundary
subregion at $I^+$ with width $\Delta w$ to an equivalent subregion
at $I^-$, stretching in the bulk time direction. For $m=0$, these
are identical to the corresponding surfaces in de Sitter discussed
in \cite{Narayan:2017xca}.
The area functional for such timelike surfaces is
\be\label{areaFnEquator}
S=\ l^{d-1} V_{S^{d-2}} \int {d\tau\over\tau^{d-1}}
   \sqrt{{1\over f(\tau)} - f(\tau) (w')^2}\ ,
\ee
where $w'={dw\over d\tau}$\,. Extremizing the area integral gives
\be
-\frac{f(\tau) w'}{\sqrt{f(\tau)^{-1}-f(\tau)w'^2}} \frac{1}{\tau ^{d-1}}=B
\quad \implies \quad
f(\tau)^2w'^2=\frac{B^2 \tau ^{2d-2}}{f(\tau)+B^2\tau ^{2d-2}}\ ,
\ee
and the extremal surfaces are given by
\be\label{wtau-B^2>0}
   {\dot w}^2 \equiv (f(\tau))^2 (w')^2
   = {B^2\tau^{2d-2}\over f(\tau)+B^2\tau^{2d-2}}\ ,
   \quad\ \ S = {2 l^{d-1} V_{S^{d-2}}\over 4G_{d+1}}
   \int_\epsilon^{\tau_*} {d\tau\over\tau^{d-1}}\
        {1\over \sqrt{f(\tau)+B^2\tau^{2d-2}}}\ .
\ee
Here ${\dot w}\equiv {dw\over dy}$ with $y$ the tortoise coordinate
$y = \int {d\tau\over f(\tau)}$ useful near the horizons: we will
discuss this further later. Requiring that ${\dot w}^2>0$ near the
boundary $\tau\ra 0$ requires $B^2>0$.

The nature of extremal surfaces as timelike or spacelike depends on the
local region containing the surface, characterized in particular by the
sign of $f(\tau)$. With $F/P$ referring to the future/past universes
in the Penrose diagram,  $I_F/I_p$ the future/past interior regions
and $D$ the diamond shaped region between the horizons, we have
{\small
\bea\label{FD-ti-sp-like}
F,P,I_F,I_P:\ \ 
f(\tau)>0 ,\ \ \ 0<\tau<{1\over a_1}\ \ {\rm or}\ \ \tau>{1\over a_2} :&&
{\dot w}^2<1\ {\rm (timelike)},\ \ {\dot w}^2>1\ {\rm (spacelike)},
\nonumber\\
D:\qquad\qquad\quad\ f(\tau)<0 ,\ \ \ {1\over a_1} < \tau < {1\over a_2} :&&
{\dot w}^2>1\ {\rm (timelike)},\ \ {\dot w}^2<1\ {\rm (spacelike)}.\quad\ \
\eea }
In other words, $w$ is a timelike coordinate in $D$, while it is a
spacelike coordinate in $F, P, I_F, I_P$.

Thus the surface (\ref{wtau-B^2>0}) satisfies ${\dot w}^2<1$ in the
future/past universes $F,P$, and so is timelike. 
Further, in the diamond $D$, we have ${\dot w}^2>1$, continuing to be
timelike. At the horizons, $f=0$ and so ${\dot w}^2=1$.

The turning point of an extremal surface such as (\ref{wtau-B^2>0}) is
the location where ${\dot w}^2\ra\infty$\,, with the surface roughly
beginning to retrace its behaviour until that point: this is typically
given by the zero of the denominator in (\ref{wtau-B^2>0}).  Since
$f>0$ in $F,P$, we have ${\dot w}^2<1$ from (\ref{wtau-B^2>0}): thus
the surface $w(\tau)$ cannot have a turning point there.  However a
turning point exists in the diamond $D$ since $f<0$ in $D$. This
structure is similar to that of the timelike surfaces in de Sitter in
\cite{Narayan:2017xca}. The turning point in this case satisfies
\be\label{tau*B^2>0}
f(\tau_*) + B^2\tau_*^4 = 0 \ ,\qquad {1\over a_1} < \tau_* < {1\over a_2}\ .
\ee
In the limit where $B\ra 0$, we see that $\tau_*$ approaches a zero
of $f$, \ie\ $\tau_*$ approaches either the cosmological
($\tau_*={1\over a_1}$) or the Schwarzschild ($\tau_*={1\over a_2}$)
horizon, so we have $B_1^2\tau_{*1}^4=f(\tau_{*1})$ and
$B_2^2\tau_{*2}^4=f(\tau_{*2})$ at first sight.

In what follows, we will find surfaces that approach the vicinity of
the cosmological horizon: these are shown as the red curves in
Figure~\ref{SdSsurf0} and are analogous to the surfaces in
\cite{Narayan:2017xca} in pure de Sitter, with $m=0$ or $a_1=1,
a_2=0$.  This begs the question of whether there are extremal surfaces
that in some limit approach the Schwarzschild horizon: on the face of
it, the turning point equation (\ref{tau*B^2>0}) suggests a distinct
branch for $\tau_*\ra {1\over a_2}$ as well. Pictorially one might
imagine surfaces represented by the dashed blue curves in
Figure~\ref{SdSsurf0}.  If such surfaces exist, one might wonder if
there are analogs of ``disentangling'' transitions observed in
holographic mutual information \cite{Headrick:2010zt}: \ie\ for given
width $\Delta w$, there would be timelike surfaces passing either near
the cosmological or the Schwarzschild horizon (akin to the solid red
or dashed blue curves respectively in Figure~\ref{SdSsurf0}), with the
ones of lower area being picked out by an area minimization
prescription.  However from Figure~\ref{SdSsurf0}, we observe that if
the dashed blue curves are timelike but approach the Schwarzschild
horizon in some limit, presumably they could only exist for
``sufficiently large'' width\ (\eg\ a ``nearly null'' timelike surface
would pass close to the Schwarzschild horizon only if it begins near
the edge of $I^+$).

\begin{figure}[h] 
\hspace{1pc}
\includegraphics[width=18pc]{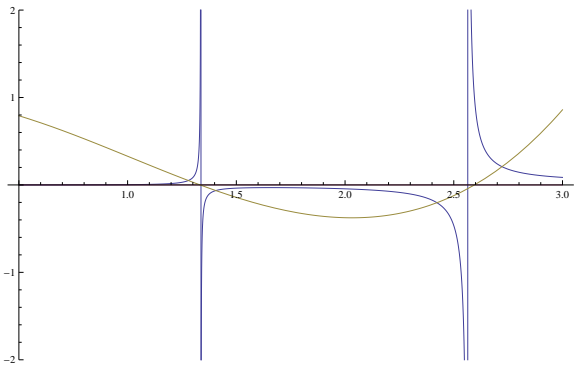} \hspace{2pc}
\begin{minipage}[b]{20pc}
  \caption{{\label{wdot2-B^2>0}\footnotesize {\ \ ${\dot w}^2$ vs $\tau$,
      for $B^2=0.001$, \newline $a_1=0.75,\ a_2\sim 0.39,\ {m\over l}\sim 0.16$.
      Here $B^2$ is \newline sufficiently small so that the turning
      points \newline where ${\dot w}^2\ra\infty$ are fairly close to,
      although \newline not precisely, the horizon values ${1\over a_1}$
      and ${1\over a_2}$\,. \newline
      Also plotted is $f(\tau)$ for these values. \newline
}}}
\end{minipage}
\end{figure}

Now we will argue that real extremal surfaces of this kind with
$B^2>0$ in fact exist only for $\tau_*$ near the cosmological
horizon. This is because ${\dot w}^2<0$ near the other branch, and more
generally between the two turning points: thus it cannot be a real
surface. To elaborate, let us recast as
\be\label{wdot2}
   {\dot w}^2 = {B^2\tau^4\over V(\tau)}\ ,\qquad V(\tau)=f(\tau)+B^2\tau^4\ ,
\ee
and $V(\tau_{*1})=0$ where $\tau_{*1}$ is the turning point near the
cosmological horizon ${1\over a_1}$\ which satisfies the range in
(\ref{tau*B^2>0}). Note that $\tau$ increases from the boundary at
$\tau=\epsilon$ to the cosmological horizon ${1\over a_1}$ and then to
the Schwarzschild horizon ${1\over a_2}$ in this coordinate
parametrization. Since this is a first order zero of $V(\tau)$,
expanding near $\tau_{*1}={1\over a_1}$ gives
\be\label{wdot2neartau1}
{\dot w}^2\ \sim\ {B^2\tau^4\over |V'(\tau_{*1})|\, (\tau_{*1}-\tau)}
\ee
where $V'(\tau_{*1})={dV\over d\tau}|_{\tau_{*1}}$ and we are dropping
the higher order terms in this approximation. We know from the explicit
form (\ref{tau*B^2>0}) that ${\dot w}^2>0$ for $\epsilon<\tau<{1\over a_1}$
and the absolute value in (\ref{wdot2neartau1}) reflects this. We now
see that for $\tau\gtrsim \tau_{*1}$ the sign of ${\dot w}^2$ becomes
negative. This continues to hold all the way till $V(\tau)$ hits the
other zero at ${1\over a_2}$\,: this can be seen numerically as in
Figure~\ref{wdot2-B^2>0}, where we have plotted ${\dot w}^2$ as a
function of $\tau$ for the representative values
$B^2=0.001,\ a_1=0.75$.\
Further we have $a_2={1\over 2}(\sqrt{4-3a_1^2}-a_1)\sim 0.39,\
{m\over l}={1\over 2} a_1(1-a_1^2) \sim 0.16$,\
noting that $f(\tau)$ factorizes as (\ref{SdS4-fma1a2}).
In other words, over ${1\over a_1}<\tau<{1\over a_2}$ we have ${\dot w}^2<0$
implying that $w(\tau)$ is not real-valued in this range for $B^2>0$.
As $B^2$ increases, the two values where ${\dot w}^2\ra\infty$ begin
to approach each other: at a critical value $B_{max}$ they coincide.
We will say more on this later.

Overall for $B^2>0$, we see that these surfaces give the red curves
in Figure~\ref{SdSsurf0}: these are similar to the $dS$ surfaces in
\cite{Narayan:2017xca}. At the horizon ${1\over a_1}$ we have
${\dot w}^2=1$. In the limit where $B^2\ra 0$, we see that
$\tau_*\sim {1\over a_1}+\delta$: expanding (\ref{tau*B^2>0}) using
(\ref{SdS4-fma1a2}), we obtain
\be\label{Bdelta}
\tau_*\sim {1\over a_1}+\delta\,:\qquad
B^2 \sim\ a_1^5 \Big( 2 - {a_2(a_1+a_2)\over a_1^2} \Big)\, \delta \equiv\
a_1^5 c \delta\ .
\ee
For $a_1=1, a_2=0$, we have $m=0$ and this becomes $B^2\sim 2\delta$
in agreement with the corresponding expression for the $dS_4$ surface
in \cite{Narayan:2017xca}. More detailed properties of these surfaces
can be obtained by using the tortoise coordinate
\bea\label{ytau}
&& y = \int {d\tau\over f(\tau)}
= \int\frac{d\tau}{(1-a_1\tau)(1-a_2\tau)(1+(a_1+a_2)\tau)}\ \nonumber\\
&&\ \ \ \xrightarrow{\,\tau\sim 1/a_1\, }\ \
- {1\over a_1 (2+{a_2\over a_1})(1-{a_2\over a_1})}\, \log |1-a_1\tau|\
\equiv\ -{1\over a_1c} \log |1-a_1\tau|\ ,
\eea
with the last approximation valid in the vicinity of the cosmological
horizon as $B^2\ra 0$\  (more detailed expressions appear in Appendix \ref{app:Penrosediag}).
For the region $\tau\gtrsim {1\over a_1}$ we have $\tau_*\ra {1\over a_1},\
y_*\ra\infty$, so using (\ref{Bdelta}), (\ref{ytau}), gives
\be\label{approx-cB}
c=(2+{a_2\over a_1})(1-{a_2\over a_1})\ ,\quad\ \
a_1\delta \sim e^{-a_1cy_*}\ ,\quad\ \ B^2\sim a_1^4c e^{-a_1cy_*}\ ,\quad\ \ 
\tau \sim {1\over a_1} (1+e^{-a_1cy})\ .
\ee
The area of these surfaces can be evaluated using (\ref{wtau-B^2>0}) and the divergent part of the area arises from the region near the boundary ($\tau\sim \epsilon\ra 0$) as
\bea\label{areaB^2>0}
S &=& {2 l^2 V_{S^1}\over 4G_4} \int_\epsilon^{\tau_*} {d\tau\over\tau^2}\
{1\over \sqrt{1-\tau^2+{2m\over l}\tau^3+B^2\tau^4}}\ ; \nonumber\\
&& S^{div} \sim {\pi l^2\over G_4} \int_\epsilon^{1/a_1} {d\tau\over\tau^2}\
{1\over \sqrt{(1-a_1\tau)(1-a_2\tau)(1+(a_1+a_2)\tau)}}\ .
\eea
This gives an area law divergence, scaling as de Sitter entropy ${\pi
  l^2\over G_4}$, not surprisingly. For $a_1=1, a_2=0$, this matches
with the $dS_4$ results in \cite{Narayan:2017xca}. As in that case,
codim-2 surfaces stretching between $I^\pm$ on a $w=const$ slice are
difficult to analyse but the area law divergence is straightforward to
see, essentially similar to (\ref{areaB^2>0}). As such, these area
integrals are certain kinds of elliptic integrals\footnote{Scaling $a_1$
  in as $x=a_1\tau$ gives\ $S \sim\ {\pi l^2 a_1\over G_4}
  \int_\epsilon^1 {dx\over x^2}\ {1\over \sqrt{(1-x)(1-\beta
      x)(1+(1+\beta)x)}}$ with $\beta={a_2\over a_1}$\,.  } and can be
evaluated numerically for any particular values of $a_1, a_2$, or
equivalently $m$. This numerical evaluation shows that the area
(\ref{areaB^2>0}) of these surfaces for Schw-$dS$ is
always greater than the corresponding area for de Sitter (keeping the
same asymptotic structure and cutoff): thus the area law divergence
scaling is bigger. In Appendix~\ref{app:sds3}, we analyse 3-dim
Schwarzschild de Sitter spacetimes and these timelike surfaces: the
area of these surfaces in $SdS_3$ can here be seen explicitly
to be greater than the corresponding area for $dS_3$\ (see
(\ref{areaSdS3dS3})).

Analytically speaking, the integral (\ref{areaB^2>0}) can be expressed
in terms of elliptic integrals and functions, as discussed in appendix
\ref{app.3}. This agrees with the de Sitter case and has a
nonvanishing expression for the physical range of $a_1, a_2$, in
(\ref{SdS4-fma1a2}). 

The finite part of the area integral (\ref{areaB^2>0}) vanishes in the
strict $B\ra 0$ limit: for an infinitesimal $\delta B^2\neq 0$ deformation
about the $B^2=0$ limit, the finite part of the area can be estimated
using the approximations (\ref{approx-cB}) as
\be
\delta S\, \sim\ {1\over a_1}\, \delta B^2\ {l^2\over G_4}\ 
\sim\ {1\over a_1}\, {l^2\over G_4}\ (2+{a_2\over a_1})(1-{a_2\over a_1})\,
a_1^5\, \Big(\tau_*-{1\over a_1}\Big)\ ,
\ee
the scaling encoding details of the SdS$_4$ mass $m$ through the parameters
$a_1,a_2$. 

As stated earlier, the quartic in the denominator of ${\dot w}^2$ in
(\ref{wdot2}) has two positive zeros for generic $B^2\gtrsim 0$. As
$B^2$ increases, these approach each other and finally coincide at
$B_{max}$ where ${\dot w}^2$ acquires a double zero in the denominator:
now $\Delta w \sim \int_0^{y_*} dy\, {\dot w}$ acquires a logarithmic
divergence and the subregion at $I^\pm$ becomes the whole space. To
see what $B_{max}$ is in greater detail, note that $B$ acquires a maximum
value at
\be\label{Bmax}
B^2 = -{f(\tau_*)\over\tau_*^4}\quad \xrightarrow{\ max\ }\quad
{dB^2\over d\tau_*}=0\ \ \Rightarrow\ \ f'(\tau_*) = {4f(\tau_*)\over\tau_*}\ .
\ee
Then expanding near $\tau=\tau_*$ gives\
\be
f(\tau)+B_{max}^2\tau^4 =
f(\tau_*)+f'(\tau_*)(\tau-\tau_*)-{f(\tau_*)\over\tau_*^4}
(\tau_*+\tau-\tau_*)^4+\ldots
\ee
which shows the constant and linear terms in $(\tau-\tau_*)$ cancelling,
using (\ref{Bmax}): this gives a double zero. For instance, $f(\tau)$ in
(\ref{SdS4-fma1a2}) gives
\bea
&&   {m\over l} = -{2-\tau_*^2\over\tau_*^3}\quad\Rightarrow\quad
   f(\tau_*) = -(3-\tau_*^2)\,,\ B_{max}^2={3-\tau_*^2\over\tau_*^4}\nonumber\\
&&   \Rightarrow\quad
   f(\tau)+B_{max}^2\tau^4 = \Big(1-{\tau\over\tau_*}\Big)^2
   \Big(1+{2\tau\over\tau_*} + (3-\tau_*^2){\tau^2\over\tau_*^2}\Big)\ ,
\eea
with a double zero at $\tau_*$, the expression factorizing. Now
$\Delta w$ acquires a logarithmic divergence as\
$\Delta w \sim {2\over \sqrt{(3-\tau_*^2)(6-\tau_*^2)}} \int^{\tau_*} {d\tau\over 1-\tau/\tau_*} \sim -{2\tau_*\over \sqrt{(3-\tau_*^2)(6-\tau_*^2)}}\log(1-{\tau\over\tau_*})$\,.
Beyond $B_{max}$ there is no real solution to $\tau_*$, giving a limiting
surface at the $\tau_*(\equiv \tau_*^{max})$ value corresponding to $B_{max}$. This is very
similar to the limiting surface in de Sitter space discussed in \cite{Narayan:2020nsc}. At this point, the finite part of the area\
scales as $S^{fin}\sim \frac{\pi \l^2}{G_4}{1\over \tau_*^2\sqrt{6-\tau_*^2} }\int_{1/a_1}^{\tau_*}
{d\tau\over 1-\tau/\tau_*} \sim {\pi\l^2\over G_4}\frac{\sqrt{3-\tau_*^2}}{2\tau_*^2}\Delta w$ with the
same logarithmic divergence, upto numerical factors.

Now we compare the limiting surface at $\tau_*^{max}$ in the Schw-$dS$ space above with the limiting surface in de Sitter space in \cite{Narayan:2020nsc}. From the expression $\frac{m}{l}=-\frac{2-\tau_*^2}{\tau_*^3}$, we see that the value of $\tau_*^{max}$ lies in the range $\sqrt{2}\leq \tau_*^{max} < \sqrt{3}$ for $0 \leq \frac{m}{l} < \frac{1}{3\sqrt{3}}$. In particular, $\tau_*^{max}=\sqrt{2}$ for $m=0$, corresponding to the limiting surface in de Sitter space. As $m$ increases, $\tau_*^{max}$ increases, showing that the limiting surface in Schw-$dS$ space lies further in the interior, compared to the limiting surface in de Sitter space.

Thus overall, we have obtained timelike extremal surfaces stretching
between $I^\pm$ passing through the vicinity of the cosmological
horizon.  For the case $m=0$, or equivalently $a_1=1, a_2=0$, we
obtain empty de Sitter space and recover the results of the timelike
extremal surfaces in \cite{Narayan:2017xca,Narayan:2020nsc}.  As we
have seen, similar surfaces passing near the Schwarzschild horizon do
not exist: if we insist on surfaces that are anchored at the future
boundary $I^+\in F$, the surfaces $w(\tau)$ above do not have the
desired behaviour for $B^2>0$ since ${\dot w}^2<0$ between the two
turning points, thus precluding real surfaces. This appears consistent
with the fact that as the subregion becomes the whole space at
$I^\pm$, the surfaces approaching the limiting surface and do not go
beyond.

 Finally, the area functional for codim-1 surfaces in $SdS_{d+1}$\ is
 $S = l^dV_{S^{d-1}} \int {d\tau\over\tau^d} \sqrt{{1\over
     f}-f(w')^2}$\,, scaling as $l^d$: these wrap the $S^{d-1}$ and
 stretch in the $(\tau,w)$-plane. The resulting extremization are
 similar structurally to (\ref{areaFnEquator}), (\ref{wtau-B^2>0}),
 except with different $\tau$-factors: as such these can be seen to be
 equivalent to codim-2 surfaces in $SdS_d$. It may be interesting to
 analyse these further.

\section{Schw-$dS$ extremal surfaces with $B^2<0$}

We are looking for surfaces that pass through the vicinity of the
Schwarzschild horizon: as we have seen, the surfaces
(\ref{wtau-B^2>0}) with $B^2>0$ do not do so. However note that the
local equation governing extremal surfaces is necessarily given by
(\ref{wtau-B^2>0}) with the conserved quantity $B$ as a
parameter. Thus the only other possibility is to check if the surfaces
(\ref{wtau-B^2>0}) with $B^2$ continued to $B^2<0$ exhibit any new
behaviour. In what follows, we will study this in detail and find that
these in fact are spacelike surfaces with interesting behaviour: they
pass through the vicinity of the Schwarzschild horizon (as well as
the cosmological one), but never reach $I^+$ as real surfaces.
Thus consider (\ref{wtau-B^2>0}) with $B^2<0$\ (focussing on $SdS_4$).
This gives the surfaces
\be\label{wtau-B^2<0}
{\dot w}^2 \equiv (f(\tau))^2 (w')^2
= {A^2\tau^4\over A^2\tau^4-f(\tau)}\ , \qquad A^2>0\ .
\ee
This can be thought of as the result of extremizing the area functional
for spacelike surfaces,
\be
S=\ l^{d-1} V_{S^{d-2}} \int {d\tau\over\tau^{d-1}}
   \sqrt{f(\tau) (w')^2 - {1\over f(\tau)}}\ .
\ee
As before, these are codim-2 surfaces: they wrap an $S^1$ in some
equatorial plane of the $S^2$ and are curves in the $(\tau,w)$-plane.
There is no interpretation for the width $\Delta w$ for these surfaces
since as we will see, these surfaces never reach the future/past
boundaries $I^\pm$. It thus appears difficult to interpret them via
$dS/CFT$.

Noting (\ref{FD-ti-sp-like}), we see that the surface (\ref{wtau-B^2<0})
satisfies ${\dot w}^2<1$ in the diamond $D$ and ${\dot w}^2>1$ in
$F,P, I_F, I_P$ and is thus a spacelike surface. At the horizons, $f=0$
and ${\dot w}^2=1$.
Since ${\dot w}^2<1$ in $D$, the surface cannot have a turning point
in $D$. However ${\dot w}^2$ can have a turning point in $F, P, I_F, I_P$,
where ${\dot w}^2\ra\infty$ given by the zero of the denominator in
(\ref{wtau-B^2<0}): we have
\be\label{trngpt-B^2<0}
A^2\tau_*^4 - f(\tau_*) = 0\ ,\qquad \tau_*\in F, P, I_F, I_P\ .
\ee
This has multiple real solutions since $f>0$. Looking near $A\ra 0$ 
suggests that there are real turning point solutions near the zeros of
$f$. In fact we obtain $\tau_*^1$ near the cosmological horizon and
also $\tau_*^2$ near the Schwarzschild one (which can be confirmed
numerically). These two turning points satisfy
$0<\tau_*^1<{1\over a_1}$ lying in $F$ and $\tau_*^2>{1\over a_2}$
lying in $I_P$, and we note also that
\be
A\ra 0\quad\Rightarrow\qquad\quad \tau_*^1\ra {1\over a_1} \qquad \& \qquad
\tau_*^2\ra {1\over a_2}\ .
\ee
The surface (\ref{wtau-B^2<0}) is real and ${\dot w}^2$ is positive
between the two turning points near $\tau_*^1 < {1\over a_1}$ and
$\tau_*^2 > {1\over a_2}$\,.\

In addition, an important point to note is that ${\dot w}^2$ exhibits a
\emph{nonzero} minimum in $D$ for $A^2\neq 0$: this minimum becomes
vanishingly small as $A^2\ra 0$. This behaviour can be seen in the plot
of (\ref{wtau-B^2<0}) against $\tau$\ (which is qualitatively similar to
Figure~\ref{wdot2-B^2>0} in the region between the turning points,
except for an overall minus sign). The significance of this nonzero
minimum is that $w(\tau)$ cannot be tangent to any $w=const$
hypersurface in the $D$ region: for if it were, then ${dw\over d\tau}$
must vanish. This implies that the surface $w(\tau)$ necessarily crosses
every $w=const$ slice in $D$, without being tangent to any slice: thus
it cannot approach the future horizon bounding $I_F$. Instead the
surface crosses the past horizon bounding $I_p$ and has a turning point
$\tau_*^2\ra {1\over a_2}$ in $I_P$ rather than $I_F$.
This curious feature of the surface $w(\tau)$ only arises due to the
specific cubic form for $f(\tau)$ and the corresponding behaviour of
${\dot w}^2$ with nonzero minimum: no such feature arises in the pure
de Sitter case, or in the timelike surfaces discussed earlier (which
have only one relevant turning point $\tau_*\ra {1\over a_1}$\,).

\begin{figure}[h] 
\hspace{5pc}
\includegraphics[width=27pc]{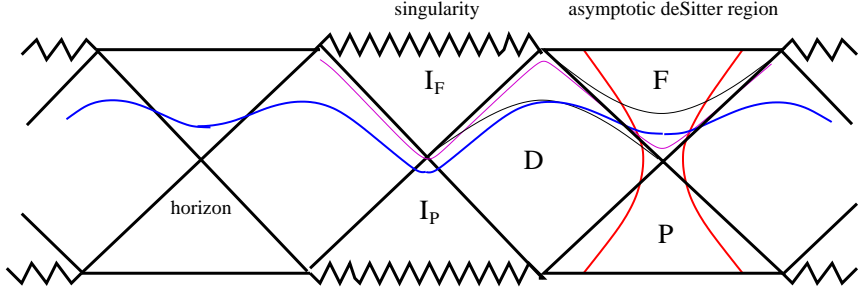} 
\caption{{\label{dSstSurfEnt}\footnotesize {Extremal surfaces
      in the $(\tau,w)$-plane of Schw $dS$: the red curve is timelike,
      from $I^+$ to $I^-$. The solid blue curve is real and spacelike
      but does not reach till $I^+$. The blue curve approaches the purple curve as $A$ decreases to a near-vanishing value.
}}}
\end{figure}

At the turning points $\tau_*^1\in F$ and $\tau_*^2\in I_P$, the
surface can be joined with similar surfaces from the other universes:
thus the surface can be extended as a spacelike surface traversing
through all the universes (stretching indefinitely). 
Overall this gives the blue curve in Figure~\ref{dSstSurfEnt}.
(Similar spacelike extremal surfaces exist with turning points
$\tau_*^1\in P$ and $\tau_*^2\in I_F$, although we have not shown them.)
As $A\ra 0$, we have $\tau_*^2\ra {1\over a_2}$ in $I_P$.
Thus within the diamond $D$, we have ${\dot w}^2\ra 0$ so that
$w(\tau)$ approaches a $w=const$ hypersurface (shown in black in $D$ 
in Figure~\ref{dSstSurfEnt}). Likewise we have
$\tau_*^1\ra {1\over a_1}$ in $F$: here ${\dot w}^2\ra\infty$
with this $\tau=const$ slice approaching $\tau={1\over a_1}$\,. Thus
the surface resembles the purple curve: in the strict $A\ra 0$ limit
the purple curve asymptotes to a zigzag spacelike curve stretching
indefinitely, grazing all the Schwarzschild and de Sitter horizons.
On the other hand, as $A$ increases from $A=0$ (which is akin to the
limiting purple curve), we find that the two zeros $\tau_*^{1,2}$ of the
quartic turning point equation (\ref{trngpt-B^2<0}) move in opposite
directions: \ie\ $\tau_*^1 < {1\over a_1}$ and $\tau_*^2 > {1\over a_2}$\,.
As in the cosmological surfaces case with $B^2>0$, in this case
there is a limiting surface with $A=A_{max}$ which is similar in shape
to the blue curve, with limiting values for $\tau_*^{1,2}$. This can
also be confirmed by numerically evaluating the zeros of (\ref{trngpt-B^2<0})
for various $A$ values. It is noteworthy that the surfaces here are
constrained with regard to two turning points which makes their
behaviour/shape different from the cosmological case. In particular,
note that the blue and purple curves cross at some nontrivial location,
since as $A$ increases from $0$ to $A_{max}$ the two zeros move
in opposite directions: in the cosmological surfaces there is a single
turning point.

These $A^2>0$ surfaces never reach $I^+$ but continue indefinitely as
spacelike surfaces, always at some distance from $I^+$.  If we require
that the surface reaches $I^+$, then it cannot be real-valued at least
(if it even exists): it must be interpreted in that vicinity as a
complex surface. For instance as $\tau\ra 0$\ (approaching $I^+$), we
have\
\be
   {\dot w}^2\sim {A^2\tau^4\over -f} \sim -A^2\tau^4 < 0\ ,
\ee
which is complex for real $\tau$, or alternatively forces an imaginary
time path $\tau=iT$ if we require real $w(\tau)$, interpreting it as
the width in the dual CFT and that it be real-valued (this sort of
argument was used in \cite{Narayan:2015vda} to
identify complex extremal surfaces in Poincare $dS$).

The area of the portion of the surface between the two turning points,
doubled, is
\be\label{areaB^2<0}
S = 2{l^2 V_{S^1}\over 4G_4} \int_{\tau_*^1}^{\tau_*^2} {d\tau\over\tau^2}\
{1\over \sqrt{A^2\tau^4-f(\tau)}}\ \ \xrightarrow{A\ra 0}\ \
{\pi l^2\over G_4} \int_{1/a_1}^{1/a_2} {d\tau\over\tau^2}\
{1\over \sqrt{-f(\tau)}}\ ,
\ee
where the overall factor of 2 stems from doubling the area so that we
cover the ``unit cell'' fully once, \ie\ going from $\tau_*^1$ to
$\tau_*^2$ and back to $\tau_*^1$.
Since $f<0$ for ${1\over a_1}<\tau<{1\over a_2}$\,, this area integral
is expected to be well-defined and finite. 
We confirm this in appendix \ref{app.3}, where the last integral in
(\ref{areaB^2<0}) is evaluated. The area integral for spacelike curves
is also shown to have a well defined extremal, or Nariai, limit: we
discuss further aspects of this extremal limit in what follows.

\subsection{The de Sitter limit}

It is interesting to note that the de Sitter limit given by
$a_1 =1,\ a_2 = 0$ for these spacelike surfaces and the associated
turning points are
\be
{\dot w}^2 = {A^2\tau^4\over A^2\tau^4-f(\tau)}\ ;\qquad\qquad
A^2\tau_*^4=f(\tau_*)=1-\tau_*^2\ ,
\ee
with a real solution for $\tau_*$ only in $F$ or $P$. The spatial
``ends'' of these surfaces, at $\tau\ra\infty$, appear ill-defined
however. Here ${\dot w}^2\ra 1$ and the surfaces end on the
North/South pole trajectories in $N/S$ but their endpoints appear
ill-defined: $\tau=\infty$ corresponds to the poles which have no
spatial extent. Pictorially these resemble the portion of the blue
curve lying in the de Sitter part of the Penrose diagram in
Figure~\ref{dSstSurfEnt}, and so bear resemblance to the Hartman,
Maldacena surfaces \cite{Hartman:2013qma} in the AdS black hole.
The present surface here can perhaps be better defined by introducing
a cutoff surface at large $\tau$ that encircles the poles.

One can formally calculate the area (\ref{areaB^2<0}) in the
$A\ra 0$ limit where the turning point approaches the horizon at
$\tau=1$.\ This gives\
\be
S\ \ra\ {\pi l^2\over G_4} \int_1^\infty {d\tau\over\tau^2}\
{1\over\sqrt{\tau^2-1}}\ \ra\ {\pi l^2\over G_4}\ .
\ee
Interestingly, this is finite and the value is precisely de Sitter
entropy.

As stated previously, these are codim-2 spacelike surfaces wrapping an
$S^1$ in some equatorial plane of the $S^2$ and stretching as curves
in the $(\tau,w)$-plane. Thus the area arises from the $S^1$ and
$\tau$-directions, and is not the $S^2$ area per se. As for all these
spacelike surfaces we have been discussing, these ones ``hang'' at
some distance from the future/past boundaries $I^\pm$, which they
never reach. It thus appears difficult to interpret them via $dS/CFT$
a priori. These are thus quite different from the connected timelike
surfaces stretching between $I^\pm$ in \cite{Narayan:2017xca}. Those
have an area law divergence\
${\pi l^2\over G_4} \int_\epsilon^1 {d\tau\over \tau^2}
{1\over\sqrt{1-\tau^2}} = {\pi l^2\over G_4} {1\over\epsilon}$\
scaling as de Sitter entropy (perhaps consistent with the leading area
law divergence expected of entanglement entropy in a CFT in a gravity
approximation).

\subsection{The extremal, or Nariai, limit}

\begin{figure}[h] 
\hspace{5pc}
\includegraphics[width=27pc]{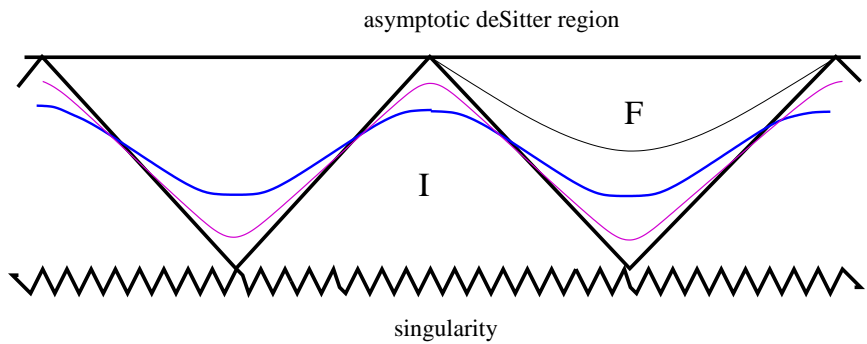} 
\caption{{\label{NariaiSurfEnt}\footnotesize {Extremal surfaces
      in the $(\tau,w)$-plane in the Nariai limit of Schw $dS$: the blue
      curve is for generic $A$, while the $A\ra 0$ limit gives the purple
      curve.
}}}
\end{figure}

As mentioned earlier, the Schwarzschild de Sitter spacetime
admits an interesting extremal, or Nariai, limit, where the values of
the cosmological and Schwarzschild horizons coincide,
\be
a_1=a_2=a_0 :\qquad f(\tau)=(1-a_0\tau)^2(1+2a_0\tau) > 0\ .
\ee
In this case, the near horizon region of the metric (\ref{SdSst2})
becomes $dS_2\times S^2$. The region $\tau\ra 0$ is the asymptotic
$dS_4$ region. In this extremal limit, we see that $f(\tau)>0$
always and thus the surfaces (\ref{wtau-B^2>0}) satisfy ${\dot w}^2<1$
and so do not exhibit any turning point (somewhat similar to the
surfaces \cite{Narayan:2015vda} in Poincare de Sitter).
However the spacelike surfaces (\ref{wtau-B^2<0}) continue to be
interesting: the turning points are
\be\label{Nariai-tau*}
A^2\tau_*^4 = (1-a_0\tau_*)^2 (1+2a_0\tau_*)\ ,
\ee
so that $\tau_*^1\leftrightarrow \tau_*^2$. Thus the blue curves now
look symmetric between the two horizons and lead to the purple curve
in the $A\ra 0$ limit (Figure~\ref{NariaiSurfEnt}).

In more detail in the Nariai limit, ${m\over l}={1\over 3\sqrt{3}}$
so that $a_0 = \frac{1}{\sqrt{3}}$ and
\begin{equation}\label{eSdS4-f}
f(\tau) = \frac{(\sqrt{3}-\tau)^2 (\sqrt{3}+2\tau)}{3\sqrt{3}}\ .
\end{equation}
The tortoise coordinate $y = \int\frac{d\tau}{f(\tau)}$ in (\ref{ytau})
becomes (see appendix \ref{app:Penrosediag})
\begin{equation}
	y =\frac{1}{\sqrt{3}-\tau} +\frac{2\sqrt{3}}{9}\log\Big|\frac{\sqrt{3}+2\tau}{\sqrt{3}-\tau}\Big| - \frac{1}{\sqrt{3}}\ .
\end{equation}
In the $A\rightarrow 0$ limit, the turning points approach the horizon
\ie\ $\tau_*^{1,2}\rightarrow \frac{1}{a_0} = \sqrt{3}$. So let us
consider the near-horizon region, $\tau\rightarrow \sqrt{3}$, where we have\
$f(\tau) \approx (\sqrt{3}-\tau)^2$\
and the tortoise coordinate can be approximated as\
$y \approx \frac{1}{\sqrt{3}-\tau}$\,.
The equation for spacelike surfaces \eqref{wtau-B^2<0} becomes
\begin{equation}
  \dot{w}^2 = \Big(\frac{dw}{dy}\Big)^2 \approx \frac{A^2(\sqrt{3}\,y-1)^4}{A^2(\sqrt{3}\,y-1)^4-y^2}\ .
\end{equation}

\medskip

\noindent \underline{In $F$ with $\tau < \sqrt{3}$}\,: The turning point
with $A>0$ and $y>0$ is given by
\begin{equation}
	A^2(\sqrt{3}\,y_* - 1)^4-y_*^2 = 0 \quad \implies \quad A(\sqrt{3}\,y_* - 1)^2=y_*\ ,
\end{equation}
whose roots are
\begin{equation}
	y_* = \frac{(2\sqrt{3}A+1)\pm \sqrt{1+4\sqrt{3}A}}{6A}\ .
\end{equation}
We see that there is a turning point $y_*^{(1)} = \frac{(2\sqrt{3}A+1) + \sqrt{1+4\sqrt{3}A}}{6A}>0$, which approaches the horizon \ie\ $y_*^{(1)} \rightarrow \infty$ as $A\rightarrow 0$.

\medskip

\noindent\underline{In $I$ with $\tau > \sqrt{3}$}\,: The turning point
with $A>0$ and $y<0$ is given by
\begin{equation}
A^2(\sqrt{3}\,y_* - 1)^4-y_*^2 = 0 \quad \implies \quad A(\sqrt{3}\,y_* - 1)^2 = -y_*\ ,
\end{equation}
whose roots are
\begin{equation}
y_* = -\frac{(1-2\sqrt{3}A)\pm \sqrt{1-4\sqrt{3}A}}{6A}\ .
\end{equation}
We see that there is a turning point $y_*^{(2)} = - \frac{(1-2\sqrt{3}A) + \sqrt{1 - 4\sqrt{3}A}}{6A}<0$ for $0 < A < \frac{1}{4\sqrt{3}}$, which also approaches the horizon \ie\ $y_*^{(2)} \rightarrow \infty$ as $A\rightarrow 0$.

Thus, we see that for spacelike surfaces described by \eqref{wtau-B^2<0}, there are turning points in both $\tau < \sqrt{3}$ and $\tau > \sqrt{3}$ regions for $0 < A < \frac{1}{4\sqrt{3}}$. These are the analogs of $\tau_*^{1,2}$
in the previous section, away from the Nariai limit. The upper bound on $A$ \ie\ $A < \frac{1}{4\sqrt{3}}$ is approximate as it is obtained by using approximate tortoise coordinate, near the horizon and for $A\rightarrow 0$. To get the exact maximum value $A_{max}$, we maximize $A(\tau_*)$ using \eqref{Nariai-tau*};
\begin{eqnarray}
	&& A^2 = \frac{(1-\frac{\tau_*}{\sqrt{3}})^2(1+\frac{2\tau_*}{\sqrt{3}})}{\tau_*^4} \quad \xrightarrow{\ max\ } \quad \frac{dA^2}{d\tau_*} = -\frac{1}{\tau_*^5}\Big(\frac{\tau_*^3}{3\sqrt{3}} -\tau_*^2 +2\Big) = 0 \nonumber \\
	&& \implies \quad A_{max}^2 = \frac{2\sqrt{3}-3}{12}\quad \text{at}\quad \tau_* = 3 + \sqrt{3}\ . \label{Amax-Nariai}
\end{eqnarray}
This $A_{max}$ leads to a limiting surface (below).

Overall this gives the spacelike extremal surfaces in the Penrose
diagram in Fig.~\ref{NariaiSurfEnt} for the surfaces (\ref{wtau-B^2<0}).
The Penrose diagram shows a region $F$ with $0\leq
\tau\leq {1\over a_0}$ containing the asymptotic de Sitter region
$\tau\ra 0$, as well as an interior region $I$ with ${1\over
  a_0}<\tau<\infty$ containing the singularity: this ``unit cell''
repeats.
As we have seen above, there are turning points in both $F$ and $I$
regions. At the horizons, we have $f(\tau)=0$ and ${\dot w}^2=1$. Away
from the horizon we have $f\neq 0$ so that
${\dot w}^2\sim {-A^2\tau^4\over f}\sim 0$ which implies that this
approaches a $w=const$ slice. Since this is true for both $F$ and $I$
regions, we obtain the purple curve in the $A\ra 0$ limit, which 
runs close to the horizons. For any infinitesimal $A\neq 0$ regulator,
the surface is spacelike, not null, and so it only approximately grazes
the horizons. As $A$ increases from $A=0$, we find that the two zeros
$\tau_*^{1,2}$ of the
quartic turning point equation \eqref{Nariai-tau*} move in opposite
directions: \ie\ $\tau_*^1 < \sqrt{3}$ and $\tau_*^2 > \sqrt{3}$\,.
As $A$ approaches $A_{max}$, there is a limiting surface similar in shape
to the blue curve, with limiting values for $\tau_*^{1,2}$.
The turning points (\ref{Nariai-tau*}) themselves can be seen to never
approach the boundary or singularity: there are no real $\tau_*$
solutions for both $\tau_*\ra 0$ (where the quartic term is negligible
relative to $f$) and $\tau_*\ra\infty$ (where the quartic term overpowers
$f$). Thus in some sense, the surfaces (\ref{wtau-B^2<0}) are repelled
from both near-boundary and near-singularity regions, corresponding to
the limiting values for $\tau_*^{1,2}$ where the limiting surface arises.

The area (\ref{areaB^2<0}) in this Nariai limit must be evaluated
carefully, by regulating $a_1=a_0-\epsilon$ and $a_2=a_0+\epsilon$.
The limits of the integral then pinch off, which may be expected to
cancel a corresponding zero from $\sqrt{-f}$ in the denominator,
thereby leading to a finite area. This is confirmed numerically. The
recasting of the area (\ref{areaB^2<0}) in terms of elliptic integrals
and functions in Appendix \ref{app.3} helps in making this precise. In
particular we obtain
\be
S_{Nariai} = {\pi^2 l^2\over 3G_4}\ ,
\ee
which is ${\pi\over 3}$ times de Sitter entropy.
The limiting surface in this case appears to run along the Schwarzschild
and cosmological horizons and the area receives contributions from
both horizons. The numerical value itself thus does not correspond
to any quantity pertaining to either horizon alone.

\subsection{Schwarzschild de Sitter and analytic continuations}

So far we have been studying the extremization problem in Schwarzschild
de Sitter directly. Now we will try to map the $SdS_4$ extremization via
analytic continuation to extremization problems in the $AdS_4$
Schwarzschild spacetime.
The Poincare version of this was discussed in \cite{Narayan:2015vda}
for the $dS_4$ black brane \cite{Das:2013mfa}, where these were
analytic continuations of the Ryu-Takayanagi expressions in the
$AdS_4$ black brane\ (in this case $f(\tau)=\tau^2-{C\over\tau}$ with
no $1$ as is usual for branes, so that the cubic admits no Nariai
limit).  The discussion below however seems slightly different since
the analytic continuation involved is distinct. Note that the
analytically continued $AdS$ cases do not appear to admit any Nariai
limit.

It is useful to recall the familiar analytic continuation
of Euclidean $AdS$ to Poincare $dS$
\be
r\ra -i\tau,\  R\ra -iR_{dS}\quad\ \Rightarrow\quad\ 
ds^2={R^2\over r^2} (dr^2+dx_i^2)\ \ \longrightarrow\ \
ds^2={R_{dS}^2\over\tau^2} (-d\tau^2+dx_i^2)\ .
\ee
In the present case, consider the 4-dim Schwarzschild de Sitter
spacetime in the form
\be\label{SdS4}
ds^2 = -f(r) dt^2 + {dr^2\over f(r)} +r^2 d\Omega_2^2\ ,
\qquad f(r)=1-{r^2\over l^2}-{2m\over r}\ .
\ee
This can be transformed to the $AdS_4$ Schwarzschild spacetime by 
analytic continuation
\be\label{AdSschw-r}
l\ra iL\qquad \Rightarrow\qquad
ds^2 = -f(r) dt^2 + {dr^2\over f(r)} +r^2 d\Omega_2^2\ ,
\qquad f(r)=1+{r^2\over L^2}-{2m\over r}\ ,
\ee
which has \emph{real} mass parameter $m$.
The boundary structure as $r\ra\infty$ is
\be
ds^2 \sim {r^2\over L^2} (-dt^2+L^2d\Omega^2) + L^2 {dr^2\over r^2}
\equiv {r^2\over L^2} (-dt^2+dx_idx_i) + L^2 {dr^2\over r^2}
\ee
where we are approximating the large $S^2$ by a plane: thus this
resembles the Poincare slicing asymptotically. For $m=0$, this simply
maps global $AdS$ to $dS_{static}$,
\be
ds^2 = -\Big(1-{r^2\over l^2}\Big) dt^2 + {dr^2\over 1-{r^2\over l^2}}
+ r^2 d\Omega_2^2\ \ \ \xrightarrow{\ l\ra iL\ }\ \ \
ds^2 = -\Big(1+{r^2\over L^2}\Big) dt^2 + {dr^2\over 1+{r^2\over L^2}}
+ r^2 d\Omega_2^2\ .
\ee
In terms of the $\tau$-coordinate, Schwarzschild de Sitter is
(\ref{SdSst}), \ie\
\be
ds^2 = {l^2\over\tau^2} \Big(-{d\tau^2\over f(\tau)} + f(\tau) dw^2
+ d\Omega_{d-1}^2\Big) , \quad\ 
f(\tau) =  1-\tau^2+{2m\over l}\tau^{2d-3}\,,\ \ \tau={l\over r}\,,\
w={t\over l}\ .
\ee
Note that $\tau, w$ are both dimensionless. It is useful to keep the
scales explicit, rewriting as
\be\label{SdS-T}
ds^2 = {l^2\over T^2} \Big(-{dT^2\over f(T)} + f(T) dW^2 + l^2d\Omega_2^2\Big),
\quad  f(T) =  1-{T^2\over l^2}+{2m\over l^4}T^3\ ,\ \
T=l\tau={l^2\over r}\,,\ w={W\over l}\,.
\ee
Now the analytic continuation above gives
\be\label{AdSschw-T}
l\ra iL,\ T\ra -T\ \ \ \longrightarrow\ \ \
ds^2 = {L^2\over T^2} \Big(- f(T) dW^2 + {dT^2\over f(T)} + L^2d\Omega_2^2\Big),
\quad  f(T) =  1+{T^2\over L^2}-{2m\over L^4}T^3\ .
\ee
We see this is identical to the earlier $AdS$ Schwarzschild metric
(\ref{AdSschw-r}) with $T={L^2\over r}$ being the effective bulk
radial coordinate. In terms of $\tau$, this analytic continuation
is\ $l\ra iL\ ,\ \tau\ra i\rho$.
Note that the AdS Schwarzschild here does not admit any analog of the
Nariai limit: this appears due to the difference in the locations of
the minus signs. Thus the Nariai limit in Schwarzschild de Sitter needs
to be treated independently, as in our discussion earlier.

The extremal surfaces (\ref{wtau-B^2>0}) in the coordinates (\ref{SdS-T})
become
\be\label{extSurf-W(T)}
{(f(T))^2\over l^2} \Big({dW\over dT}\Big)^2
= {B^2T^4/l^4\over f(T) + B^2T^4/l^4}\ . 
\ee
Under the analytic continuation (\ref{AdSschw-T}) above, these surfaces
become
\be\label{extSurf-anaCont-W(T)}
-{(f(T))^2\over L^2} \Big({dW\over dT}\Big)^2 =
{-A^2T^4/L^4\over f(T)  - A^2T^4/L^4}\ \qquad\qquad [B^2=-A^2]\ ,
\ee
where we have taken $B^2\ra -A^2$, as for the surfaces (\ref{wtau-B^2<0}).
The overall minus signs cancel giving real surfaces $W(T)$.
This is the analog of the Ryu-Takayanagi extremization in the above
$AdS$ background, but on a ``time''-slice obtained by taking one of
the equatorial planes of $S^2$.
Now in the analytically continued $AdS$-Schw case above, the turning
point $T_*$ is given by $f(T_*)-A^2T_*^4=0$ and the
width $\Delta l$ scales as $\Delta l\sim {1\over \sqrt{A}}$ in analogy
with the $AdS$ case. When $A\ra 0$, it appears that the turning
point approaches $f\ra 0$ \ie\ $T_*\ra T_h$, in other words the extremal
surface wraps the horizon (analogous to the AdS black hole case).

Thus these RT/HRT surfaces in the $AdS$ Schwarzschild under the above
analytic continuation map to the spacelike surfaces (\ref{wtau-B^2<0})
in Schwarzschild de Sitter. Note by comparison that the surfaces
(\ref{wtau-B^2>0}) in Schw-dS or dS-static \cite{Narayan:2017xca}
are not analytic continuations of any obvious $AdS$ RT/HRT extremization,
although they are analogous to rotated versions of the Hartman, Maldacena
surfaces \cite{Hartman:2013qma}.
Note also that this is all with \emph{real} mass parameter for the AdS
Schwarzschild, and so appear distinct from the analytic continuations
in \cite{Anninos:2012ft}, \cite{Maldacena:2019cbz}, arising in the
no-boundary proposal. Our goal here is to simply map the extremal
surfaces we have discussed to \emph{some} extremization problems in
$AdS$.

\medskip

\noindent {\bf $EAdS_4$ Schw\ $\ra$\ $SdS_4^{im}$ analytic continuation:}\ \ 
Consider the Euclidean $AdS_4$ Schwarzschild black hole, which is
(\ref{AdSschw-r}) with $t\ra -it$:\ the analytic continuation $r\ra -i\tau$
gives
\bea
&& ds^2 = f(r)dt^2 + {dr^2\over f(r)} + r^2d\Omega_2^2\ ,\qquad
f(r) = 1 + {r^2\over l^2} - {2m\over r} \label{SAdSm}\nn\\
\xrightarrow{r\ra -i\tau} &&
ds^2 = - {d\tau^2\over 1 - {\tau^2\over l^2} - {2mi\over\tau}} 
+ \Big(1 - {\tau^2\over l^2} - {2mi\over\tau}\Big) dt^2
- {\tau^2\over l^2} (l^2d\Omega_2^2)\ .\label{SdS-im}
\eea
For large $\tau$, this is approximated as
\be
ds^2 = -\left( -{d\tau^2\over {\tau^2\over l^2} + {2mi\over\tau}}
  + \Big( {\tau^2\over l^2} + {2mi\over\tau}\Big) dt^2
  + {\tau^2\over l^2} (l^2d\Omega_2^2)\right) .
\ee
For $m=0$, this becomes\
$l^2 {d\tau^2\over\tau^2} - {\tau^2\over l^2} (dt^2 + l^2d\Omega_2^2)$.
For large $\tau$, approximating the boundary $(t,\Omega_2)$ as
$(t,x_i)\equiv R^3$ suggests the analytic continuation $l\ra -il$ to
obtain $dS_4$, the overall minus sign being absorbed by this analytic
continuation in $l$. This approximation for nonzero $m$ gives the $dS_4$
black brane \cite{Das:2013mfa}
\be
\xrightarrow{l\ra -il} \qquad
ds^2 = -{d\tau^2\over {\tau^2\over l^2} - {2mi\over\tau}}
  + \Big( {\tau^2\over l^2} - {2mi\over\tau}\Big) dt^2
  + {\tau^2\over l^2} dx_i^2\ .
\ee
Note that in obtaining this, we have pulled a factor of $l^2$ into the
sphere and not continued that: this amounts to approximating the large
sphere of size $l$ by a plane. The final complex asymptotically $dS_4$
space can in fact be obtained via analytic continuation from the
$EAdS_4$ Schwarzschild black brane
$ds^2 = dr^2/({r^2\over R^2}-{2m\over r}) 
+ ({r^2\over R^2}-{2m\over r}) dt^2 + {r^2\over R^2} dx_i^2$.

The Ryu-Takayanagi surface on a $t=const$ slice of $EAdS_4$ Schwarzschild
can be mapped to a corresponding surface in this $dS_4$ black brane by
analytic continuation \cite{Narayan:2015vda}. The corresponding surface
is parametrized along imaginary time paths: the IR limit where the
RT-$AdS$ surface wraps the Schwarzschild horizon maps to an ``IR'' limit
corresponding to $\tau=i\tau_0$ where $\tau_0$ maps to the $AdS$
Schwarzschild horizon.
Similarly, other surfaces in the $dS_4$ brane with imaginary
mass can be obtained by analytic continuation from the $EAdS$
black brane.

This suggests that similar analytic continuations can be used to map
RT surfaces in the $EAdS$ Schwarzschild black hole (\ref{SAdSm}) to
those in the Schwarzschild de Sitter space with imaginary mass
(\ref{SdS-im})\ (with the $l^2d\Omega^2$ held fixed under analytic
continuation). It would be interesting to understand these structures,
towards understanding imaginary mass $SdS_4$ spacetimes.

\section{Comments}

We have studied extremal surfaces in the Schwarzschild de Sitter black
hole spacetime $SdS_{d+1}$ with real mass. We find codim-2 timelike
surfaces on $S^{d-1}$ equatorial planes stretching from $I^+$ to $I^-$
passing through the vicinity of the cosmological horizon: these are
analogs of the future-past surfaces in de Sitter space discussed in
\cite{Narayan:2017xca,Narayan:2020nsc}, described by ${\dot w}^2 =
{B^2\tau^{2d-2}\over f(\tau) + B^2\tau^{2d-2}}$ with $B^2>0$. There is
a limiting surface when the subregion at $I^\pm$ becomes the whole
space, as in the de Sitter case. For $B^2<0$, these are spacelike 
surfaces passing through the vicinity of the Schwarzschild and
cosmological horizons in a certain limit: these also admit an
interesting Nariai limit. These spacelike surfaces can be obtained
as certain analytic continuations of RT/HRT surfaces in $AdS$
Schwarzschild as we have seen: other analytic continuations of
(Euclidean) $AdS_4$ Schwarzschild are related to Schwarzschild de
Sitter with imaginary mass (related discussions appear for the $dS_4$
black brane \cite{Das:2013mfa}). These latter imaginary mass $SdS_4$
spaces arise in certain limits of $dS/CFT$ \cite{Anninos:2012ft}.

The future-past extremal surfaces in de Sitter are perhaps best
regarded as encoding some sort of bulk entanglement entropy, as
discussed in \cite{Narayan:2020nsc}. From this point of view, although
real mass $SdS$ spaces do not appear to arise in $dS/CFT$, one might
consider the black hole as an excitation over de Sitter and thereby
expect the area of the cosmological extremal surfaces we have
discussed to encode information about $SdS$ spaces. In this regard, we
have seen that the area of these surfaces is greater than the
corresponding area in de Sitter: as we have seen,\ $S_{SdS}\geq
S_{dS}$ in all cases for the cosmological surfaces (see the
discussions after (\ref{areaB^2>0}) for $SdS_4$ and Appendix
\ref{app:sds3} for $SdS_3$). This is reminiscent of the behaviour of
extremal surfaces in the $AdS$ black hole (where in the IR limit, the
surface hugs the horizon and the finite part of the area is the
thermal entropy \cite{Ryu:2006bv,Ryu:2006ef}): here also the area of
the surfaces increases $S_{SdS}$ with the mass perturbation to empty
de Sitter.  However this behaviour is markedly different from the
thermodynamic entropy which is known to decrease upon adding mass to
de Sitter \cite{Spradlin:2001pw}! Thus while the extremal surfaces do
not directly encode the entropy, they do scale as de Sitter entropy
and also encode the mass parameter (and thereby the entropy, if only
indirectly).

We now make a few comments on the cosmological surfaces, thinking of
them as $SdS$ generalizations of the future-past surfaces in de Sitter
space \cite{Narayan:2017xca,Narayan:2020nsc}. In particular in
\cite{Narayan:2020nsc}, a codim-1 envelope surface was discussed
arising from the family of codim-2 surfaces. This leads to an analog
of the entanglement wedge
\cite{Czech:2012bh,Wall:2012uf,Headrick:2014cta} as the codim-0
interior of the bulk region bounded by the extremal surface and the
subregions at $I^\pm$. These exhibit various geometric properties:
\eg\ the extremal surfaces lie in the causal shadow of the boundary
subregions, consistent with causality for entanglement.  Further they
also lead to an analog of subregion duality in de Sitter space: a
boundary subregion maps to the corresponding bulk subregion defined by
the entanglement wedge for the corresponding future-past extremal
surfaces.  Similar arguments can be made for the cosmological surfaces
in the present case: since real mass Schwarzschild de Sitter does not
appear to have clear interpretation in terms of $dS/CFT$, these
arguments are heuristic here but do reveal similar geometric
properties for the surfaces.  For instance, a codim-1 envelope surface
arises from the family of codim-2 surfaces which then defines an
analog of the ``entanglement wedge'' as the interior of the region
bounded by the red extremal surface curve (Figure~\ref{SdSsurf0}) and
the boundary subregions at $I^\pm$: this is entirely confined to the
$F, P, D$ regions of the spacetime. Since these surfaces do not
approach the vicinity of the Schwarzschild horizon (recalling the
existence of the limiting surface), the ``entanglement wedge'' also
does not access the vicinity of the Schwarzschild horizon. This is
perhaps consistent with the geometric fact (and causality) that the
Schwarzschild region (in particular its interior) is spacelike
separated from all points of $I^\pm$.  It is amusing to note that the
entanglement wedge region for $SdS$ is ``bigger'', in a sense, than
that in pure de Sitter. For instance the corresponding limiting
surface dips further into the interior (the $D$ region), with the
corresponding turning point value $\tau_*^{max}$ larger than that in
$dS$:\ from the discussion around (\ref{Bmax}), the limiting surface
satisfies $\sqrt{2}\leq\tau_*^{max} < \sqrt{3}$, with $\sqrt{2}$ for
$m=0$ \ie\ $dS$ and approaching $\sqrt{3}$ as $m$ approaches the
Nariai value ${m\over l}={1\over 3\sqrt{3}}$\ (in this regard, note
that the black hole moves the cosmological horizon to $\tau_1={1\over
  a_1}$\,, outward compared with de Sitter). This is perhaps
consistent with the area $S_{SdS}$ being bigger as well, as stated
above.

Overall, we note that the extremal surfaces we have been discussing
are perhaps best regarded as new probes of such cosmological
spacetimes, and would be interesting to understand better.

\vspace{7mm}

{\footnotesize \noindent {\bf Acknowledgements:}\ \ It is a pleasure
  to thank Shiraz Minwalla and Sandip Trivedi for an early discussion,
  and Juan Maldacena and Kyriakos Papadodimas for discussions as this
  work was nearing completion. KN thanks the String Theory Group,
  TIFR, Mumbai, the organizers of the Simons Summer Workshop in
  Mathematics and Physics 2019, ``Cosmology and String Theory'',
  Simons Center for Geometry and Physics, Stony Brook, USA, and the
  High Energy, Cosmology and Astroparticle Physics section, ICTP,
  Trieste, Italy for hospitality while this work was in progress.  KF
  was partially supported by the Max Planck Partner Group ``Quantum
  Black Holes'' between CMI Chennai and AEI Potsdam. The research of KK is partially supported by the SERB grant ECR/2017/000873. This work is
  partially supported by a grant to CMI from the Infosys Foundation.
}

\vspace{4mm}

\appendix

\section{Tortoise coordinates and Penrose diagrams}\label{app:Penrosediag}
In this appendix, we give constructions of the Penrose diagrams for $SdS_4$ in Fig.~\ref{dSstSurfEnt} and extremal $SdS_4$ in Fig.~\ref{NariaiSurfEnt}.

\noindent {\bf Penrose diagram for $SdS_4$}\\
For $SdS_4$ with $f(\tau)$ in \eqref{SdS4-fma1a2}, the tortoise coordinate $y=\int \frac{d\tau}{f(\tau)}$ in \eqref{ytau} can be integrated in the $D$ region to get \cite{GuvenNunez}
\begin{eqnarray}
	&& y = -\beta_1\log(1-a_1\tau) +\beta_2\log(1-a_2\tau) +\beta_3\log(1+(a_1 + a_2)\tau)\ ; \nonumber \\[2pt]
	&& \beta_1 = \frac{a_1}{3a_1^2 -1}\ , \quad \beta_2 = -\frac{a_2}{3a_2^2 -1}\ , \quad \beta_3 = \frac{a_1 + a_2}{3a_1a_2 +2}\ , \qquad \beta_1,\beta_2,\beta_3 > 0\ .
\end{eqnarray}
In the $D$ region, $y\rightarrow \infty$ as $\tau\rightarrow \tau_d=\frac{1}{a_1}$ and $y\rightarrow -\infty$ as $\tau\rightarrow \tau_s=\frac{1}{a_2}$. The ingoing and outgoing radial null coordinates are $(w-y)$ and $(w+y)$ respectively and using these we can define Kruskal-type coordinates. However, the entire Penrose diagram for $SdS_4$ cannot be covered by a single set of Kruskal-type coordinates. We use two sets of coordinates\,: the Kruskal-type coordinates, $U_2 = -e^{-\frac{w-y}{2\beta_2}}$ and $V_2 = e^{\frac{w+y}{2\beta_2}}$ around the Schwarzschild horizon (\ie\ in $D$ and $I$ regions) and Gibbons-Hawking coordinates, $U_1 = e^{\frac{w-y}{2\beta_1}}$ and $V_1 = -e^{-\frac{w+y}{2\beta_1}}$ around the cosmological horizon (\ie\ in $F$ and $D$ regions) \cite{GuvenNunez}. From $U_1 V_1 = -e^{-\frac{y}{\beta_1}} \rightarrow -\infty$ as $y\rightarrow -\infty$ at $\tau=\frac{1}{a_2}$, we see the coordinates $(U_1,V_1)$ break down at the Schwarzschild horizon. Similarly, $U_2 V_2 = -e^{\frac{y}{\beta_1}} \rightarrow \infty$ as $y\rightarrow \infty$ at $\tau=\frac{1}{a_1}$ showing that $(U_2,V_2)$ break down at the cosmological horizon.

As $(w-y)$ varies from $-\infty$ to $\infty$, $U_1$ varies from $0$ to $\infty$, and as $(w+y)$ varies from $-\infty$ to $\infty$, $V_1$ varies from $-\infty$ to $0$. Extending the range of $U_1$, $V_1$ to $U_1\in (-\infty,\infty)$ and $V_1\in (-\infty,\infty)$, then covers the $4$ regions\,: $F$, $P$ and two adjacent $D$ (which include the red curves) in Fig.~\ref{dSstSurfEnt}. Similarly, as $(w-y)$ varies from $-\infty$ to $\infty$, $U_2$ varies from $-\infty$ to $0$, and as $(w+y)$ varies from $-\infty$ to $\infty$, $V_2$ varies from $0$ to $\infty$. Then extending their range to $U_2\in (-\infty,\infty)$ and $V_2\in (-\infty,\infty)$ covers the $4$ regions\,: $I_F$, $I_P$ and two adjacent $D$ (which include the blue curve) in Fig.~\ref{dSstSurfEnt}.

\noindent {\bf Penrose diagram for extremal $SdS_4$}\\
For the extremal $SdS_4$ with $f(\tau)$ in \eqref{eSdS4-f}, the tortoise coordinate $y=\int\frac{d\tau}{f(\tau)}$ can be integrated to get
\begin{equation}
	y = \frac{1}{\sqrt{3}-\tau} +\frac{2\sqrt{3}}{9}\log\Big|\frac{\sqrt{3}+2\tau}{\sqrt{3}-\tau}\Big| - \frac{1}{\sqrt{3}}\ ,
\end{equation}
which is normalized so that at the future boundary $I^+$ at $\tau = 0$, $y = 0$. As we approach the horizon in $F$ and $I$ regions \ie\ $\tau\rightarrow \tau_0^-$ and $\tau\rightarrow \tau_0^+$, $y\rightarrow \infty$ and $y\rightarrow -\infty$ respectively. At the singularity $\tau\rightarrow \infty$, $y\equiv y_c = \frac{2\sqrt{3}}{9}\log 2 -\frac{1}{\sqrt{3}} < 0$. The discontinuity in tortoise coordinate $y$ at the degenerate horizon $\tau_0=\sqrt{3}$ suggests that we use different sets of coordinates in the $F$ and $I$ regions.

The Penrose diagram Fig.~\ref{NariaiSurfEnt} is for extremal white-holes and following \cite{Podolsky:1999ts}, we define the Kruskal-type coordinates $(\tilde{u},\tilde{v})$ as $u=y_c \cot\tilde{u}$ and $v=y_c\tan\tilde{v}$. Here $u=w-y$ and $v=w+y$ are the ingoing and outgoing null coordinates respectively. In Fig.~\ref{NariaiSurfEnt}, the $\tilde{u}$ and $\tilde{v}$ axes are at angles $\frac{3\pi}{4}$ and $\frac{\pi}{4}$ with respect to $I^+$. In the $F$ region, the left and right horizons correspond to $v\rightarrow\infty$, $\tilde{v}=-\frac{\pi}{2}$ and $u\rightarrow -\infty$, $\tilde{u}=0$ respectively. In the $I$ region, the left and right horizons correspond to $u\rightarrow\infty$, $\tilde{u}=0$ and $v\rightarrow -\infty$, $\tilde{v}=\frac{\pi}{2}$ respectively.

\section{$SdS_3$ and extremal surfaces}\label{app:sds3}

The 3-dim case is somewhat special so we analyze it here separately. The 
metric (\ref{SdSst}) for $SdS_3$ is
\be\label{ns.sds3}
ds^2 = - f(r) dt^2 + \frac{1}{f(r)} dr^2 + r^2 d\phi^2\,,\qquad 
f(r) = 1 - 8G_3E - \frac{r^2}{l^2}\ .
\ee
Unlike in higher dimensions, there is only one root here where $f(r)=0$,
namely 
\begin{equation}
r_C = l \alpha\,;\qquad \alpha = \sqrt{1 - 8 G_3 E}\ .
\label{hor.sds3}
\end{equation}
Thus the spacetime admits only one horizon, as in de Sitter.  This
metric (\ref{ns.sds3}) describes an asymptotically de Sitter spacetime
with a pointlike object of mass $E$.  When $E=0$, we have $\alpha = 1$
and we recover $dS_3$.  On the other hand, $E = \frac{1}{8G_3}$ gives
$\alpha=0$: here the horizon pinches off giving a degenerate limit.

Recasting in terms of the $\tau$-coordinate as in (\ref{SdSst2}) gives
\begin{equation}
ds^2 = \frac{l^2}{\tau^2} \left( -\frac{d\tau^2}{1-\alpha^2\tau^2}
  + (1 - \alpha^2\tau^2) dw^2 + d\phi^2\right) .
\label{fp.sds3}
\end{equation}
Restricting to the equatorial plane ($\phi = \frac{\pi}{2}$) and simplifying
the area functional (\ref{areaFnEquator}) leads to timelike extremal
surfaces (\ref{wtau-B^2>0}), which become
\be\label{wdotareaSdS3}
{\dot w}^2 \equiv (f(\tau))^2 (w')^2
   = {B^2\tau^{2}\over 1-\al^2\tau^2+B^2\tau^2}\ ,
   \quad\ \ S = {l\over G_3}
   \int_\epsilon^{\tau_*} {d\tau\over\tau}\
        {1\over \sqrt{1-\al^2\tau^2+B^2\tau^2}}\ .
\ee
Here ${\dot w}$ refers to the $y$-derivative with $y$ the tortoise
coordinate. The turning point $\tau_*$ where $\dot{w}\to\infty$ is, from
(\ref{wdotareaSdS3}),
\begin{equation}
  1 - (\alpha^2 - B^2) \tau_*^2 = 0\ , \quad \ie \quad
  \tau_* = \frac{1}{\sqrt{\alpha^2 - B^2}}\ .
\label{tp.sds3}
\end{equation}
Thus a real turning point exists only if $B^2 \leq \alpha^2$. For any
nonzero $B$, the turning point satisfies $\tau_*>{1\over\al}$\,.
As $B\ra 0$, we have $\tau_*\ra {1\over\al}$ so that the turning point
approaches the horizon, while $\tau_*\ra\infty$ as $B\ra\al$. The area of the surface (\ref{wdotareaSdS3}) becomes
\be
S = {l\over G_3}\, \log {2\tau_*\over\epsilon}\ \sim\
{l\over G_3}\, \log {2/\sqrt{1-8G_3E}\over\epsilon}\ .
\ee
To compare this with the corresponding area in $dS_3$, we obtain
\be\label{areaSdS3dS3}
S_{SdS_3}-S_{dS_3} = {l\over G_3} \log {1\over\sqrt{1-8G_3E}}\ ,
\ee
where we are comparing with the same asymptotic structure and cutoff.
We see that the area is larger for the $SdS_3$ case.

The tortoise coordinate in this case has a simple form
\begin{equation}
y = \int \frac{d \tau}{1 - \alpha^2 \tau^2} = \frac{1}{2 \alpha}\, \log\left \vert \frac{1 + \alpha \tau}{1 - \alpha \tau} \right \vert\ .
\label{tort.sds3}
\end{equation}
Inverting gives
\be
\tau = \frac{1}{\alpha}\,\frac{e^{2y\alpha} + 1}{e^{2y\alpha} - 1} \qquad
\left[\tau > \frac{1}{\alpha}\right];\qquad
\tau = \frac{1}{\alpha}\,\frac{e^{2y\alpha} - 1}{e^{2y\alpha} + 1} \qquad
  \left[\tau < \frac{1}{\alpha}\right]\ .
\label{invtort.sds3}
\ee
To analyse the width in more detail, rewriting (\ref{wdotareaSdS3}) as\
$\dot{w}^2 = \frac{1}{1 + \frac{1 - \alpha^2 \tau^2}{B^2 \tau^2}}$\ gives
\be
\dot{w}^2 = \frac{1}{1 - \frac{4 \alpha^2 e^{2y\alpha}}{B^2 \left(e^{2y\alpha} + 1\right)^2}}  \qquad\ \left[\tau > \frac{1}{\alpha}\right]\, ;\qquad\qquad
\dot{w}^2 = \frac{1}{1 + \frac{4 \alpha^2 e^{2y\alpha}}{B^2 \left(e^{2y\alpha} - 1\right)^2}}  \qquad\  \left[\tau < \frac{1}{\alpha}\right]
\label{wdot2.sds3}
\ee
The turning point relation (\ref{tp.sds3}), using (\ref{invtort.sds3}),
can be written as
\be\label{By*SdS3}
B^2 = \al^2-{1\over\tau_*^2} = {4\al^2e^{2\al y_*} \over (e^{2\al y_*}+1)^2}\ .
\ee
This can be used in (\ref{wdot2.sds3}) to estimate the width scaling.
The full width integral can be written as\
$\Delta w = 2\int_0^{y_*} dy\, {\dot w}
= 2\int_0^Y dy\, {\dot w} + 2\int_Y^{y_*} dy\, {\dot w}$ where $Y\ra\infty$
is a cutoff near the horizon $\tau={1\over\al}$\,. Then as in
\cite{Narayan:2020nsc}, the near horizon contributions are smooth and
we obtain
\be
\Delta w \sim \log (\al^2-B^2) \sim \log \tau_*
\ee
so that $\Delta w\ra 0$ as $B\ra 0$ and $|\Delta w|\ra \infty$ as $B\ra\al$.
In this case, the finite part of the area also scales as $\log\tau_*$ thus scaling linearly
with the subregion size.

\section{Analytic expressions for the area integrals} \label{app.3}

In the text, we have discussed the timelike and spacelike extremal
surfaces in the corresponding cases and obtained their areas,
respectively in (\ref{areaB^2>0}) and (\ref{areaB^2<0}).  These 
integrals in (\ref{areaB^2>0}) and (\ref{areaB^2<0}) can in fact be
analytically characterized in terms of  elliptic integrals and functions.
We here mention key definitions and identities needed to discuss the
results following the conventions of \cite{BF:1971}, which we refer to
for further details. The usual elliptic integrals of the first and
second kind are defined by
\begin{align}
F(\phi,k) &= \int \limits_{0}^{\phi} \frac{d \theta}{ \sqrt{1- k^2 \, \sin^2\theta}} = \int \limits_{0}^{\sin\,\phi}  \frac{d t}{ \sqrt{(1- t^2)(1- k^2 t^2)}}\ , \notag\\
E(\phi,k) &= \int \limits_{0}^{\phi} \sqrt{1- k^2\, \sin^2\theta} d \theta = \int \limits_{0}^{\sin\,\phi}  \sqrt{\frac{1- k^2 t^2}{1- t^2}} dt\ .
\label{el.int}
\end{align}
The above integrals are functions of a modulus $k$ and an argument, $y = \sin \phi$, where $0 \le y \le 1$, or $0 \le \phi \le \frac{\pi}{2}$. If $\phi = \frac{\pi}{2}$ (or $y=1$), then the above integrals are said to be complete and are denoted by
\begin{equation}
 K(k) = F\left(\frac{\pi}{2},k\right) \,, \quad  E(k) = E\left(\frac{\pi}{2},k\right) \,.
\end{equation}
Further they can be expressed in terms of hypergeometric functions as\
$K(k) = \frac{\pi}{2} \,_2F_1(\frac{1}{2};\frac{1}{2};1;k^2)$\ and\
$E(k) = \frac{\pi}{2} \,_2F_1(-\frac{1}{2};\frac{1}{2};1;k^2)$\  when
$k^2<1$.

For all other $\phi$, the elliptic integrals are incomplete. The
incomplete elliptic integrals will be denoted by $F(v)$ and $E(v)$,
where $v$ denotes $\{\phi,k\}$. The above elliptic integrals of
the first and second kind are real valued when $0 \le k \le 1$.

The expressions involved in the context of elliptic integrals are often
described in terms
of certain Jacobi elliptic functions. For our purposes, we will be
concerned with the elliptic functions $\text{sn} (v)\,, \text{cn} (v)$
and $\text{dn} (v)$, which satisfy
\begin{equation}
  \sin \phi \equiv \text{sn} (v) \,,\quad
  \text{cn} (v) = \sqrt{1 - \text{sn}^2(v)} \,,
  \quad \text{dn} (v) = \sqrt{1 - k^2 \text{sn}^2(v)}  \ .
\label{el.fn}
\end{equation}
\begin{equation}
  \text{sn}^2(v)+\text{cn}^2(v) =1\, \qquad  1-k^2 \text{sn}^2(v)
  =\text{dn}^2(v)\, \qquad \frac{d}{dv}(\text{sn}(v))
  =\text{cn}(v)\text{dn}(v)\ . \label{jac.id}
\end{equation}
We will evaluate the integrals in (\ref{areaB^2>0}) and (\ref{areaB^2<0})
by performing a transformation which casts the integral in a known form
involving these elliptic functions. In general, integrals involving
square roots of cubic or quartic expressions of a variable $\tau$ can
be simplified by performing a general Mobius transformation\
$\tau \ra {a+bu\over c+du}$\,. This transformation shifts the roots of
the original cubic, which can thus be used to remove specific coefficients,
such as that of the cubic term etc. The limits in the new integral
follow from the inverse transformation\ $u={a-c\tau\over d\tau-b}$\ which
is used following the final change in coordinates. Specific simplifications
arise \eg\ for the cubic case, where we retain only the quadratic and
linear terms (or for a quartic, retaining only the quadratic and quartic
terms). Additional useful simplifications occur if we ensure well-defined
limits in the resulting integral as well as \eg\ $0\leq k^2\leq 1$.
If we obtain \eg\ limits such that $u\in [0,1]$, then we end up with
``definite'' elliptic integrals. More generally, we change variables to
one of the Jacobi elliptic functions above, which leads to simplifications.
Once an integral is recast in sufficiently simple form, it may be
possible to look it up in \eg\ \cite{BF:1971}.

\noindent {\bf Evaluation of the integral (\ref{areaB^2>0})}

In the case of \ref{areaB^2>0}, let us first define
\begin{equation}
\beta = \frac{a_2}{a_1} \,, \quad k^2 = \frac{\beta \left(2+\beta\right)}{1+ 2\beta} \,, \quad \alpha^2 = \frac{k^2}{\beta}\ .
\end{equation}
We then find that substituting
\begin{equation}
\tau = \frac{\frac{1}{a_1} - \frac{k^2}{a_2} \text{sn}^2 v}{1 - k^2 \text{sn}^2 v}\,,
\end{equation}
in \ref{areaB^2>0} provides the following integral
\begin{equation}
S =  \frac{\pi l^2}{G_4} \frac{2 a_1}{\sqrt{1+2\beta}}\int \limits_{0}^{F\left(\bar{v}\right)} dv\frac{\text{dn}^4 (v)}{(1-\alpha^2\text{sn}^2(v))^2}\ ,
\label{ei.t}
\end{equation}
where $\bar{v}$ in upper limit $F\left(\bar{v}\right)$ of \ref{ei.t} corresponds to
\begin{equation}
\text{sn}(\bar{v}) = \frac{1}{\alpha} \sqrt{\frac{1- a_1 \epsilon}{1 - a_2 \epsilon}}\ .
\label{t.ang}
\end{equation}
The integral in \ref{ei.t} can be expressed in terms elliptic integrals and functions (c.f. Eq. 339 of \cite{BF:1971}). Upon considering the limits of the integration, we find the following result
\begin{equation}
S =  \frac{\pi l^2}{G_4} \frac{a_1}{\sqrt{1+2\beta}}\left[(1+\beta)F(\bar{v}) - \left(1+2\beta\right) E(\bar{v}) + \left(2+\beta\right) \frac{\text{sn}(\bar{v}) \,\text{cn}(\bar{v})\,\text{dn}(\bar{v})}{1 - \alpha^2 \text{sn}^2(\bar{v})}\right].
\label{t.int}
\end{equation}
This is the exact result for the timelike surfaces (\ref{areaB^2>0}) in
Schwarzschild de Sitter. To verify the de Sitter limit $a_1=1, a_2 =0$,\ 
note that we now have\
$\beta \to 0 ,\ k \to 0 ,\ \alpha^2 \to 2$. We also have\
$F(\bar{v}) = \phi = E(\bar{v})$ and\
$\text{sn}(\bar{v}) \to \sin(\bar{v}) ,\ \text{cn}(\bar{v}) \to \cos(\bar{v}) ,
\  \text{dn}(\bar{v}) \to 1$.\ Then (\ref{t.int}) simplifies to
\begin{align}
S \to S_{dS} &= \frac{\pi l^2}{G_4}\frac{2 \sin(\bar{v}) \,\cos(\bar{v})}{1 - 2 \sin^2(\bar{v})} = \frac{\pi l^2}{G_4} \tan(2 \bar{v})\ .
\end{align}
From (\ref{t.ang}) we have $\sin\bar{v} = \sqrt{\frac{1 - \epsilon}{2}}$\
giving
\begin{equation}
S_{dS} =  \frac{\pi l^2}{G_4}\frac{1}{\epsilon} + \mathcal{O}\left(\epsilon\right)\,,
\end{equation}
in agreement with the area in de Sitter \cite{Narayan:2017xca}.

For $m \neq 0$, i.e. $a_1 > a_2 \ge 0$, we have results for the
general Schwarzschild de Sitter case. With the Schwarzschild horizon
present, $\frac{\pi}{2} - \frac{\epsilon}{2}(a_1 - a_2) \ge \bar{v}
\ge \frac{\pi}{4} - \frac{\epsilon}{2}(a_1 - a_2)$ and we can always
expand $F(\bar{v})$, $E(\bar{v})$ and the elliptic functions
$\text{sn}(\bar{v})\,,\text{cn}(\bar{v})$ and $\text{dn}(\bar{v})$ as
a power series (c.f. Eqs. 902 and 903 of \cite{BF:1971}). The values
of $k^2$ and $\alpha^2$ depend on the choice of $\frac{a_2}{a_1}$,
\ie\ $\beta$.

\medskip

\noindent {\bf Evaluation of the integral (\ref{areaB^2<0})}

To evaluate the area integral in the spacelike case, we now define
\begin{equation}
  k^2 = \frac{\left(1 - \beta^2\right)}{1+ 2\beta} \,, \quad
  \alpha^2 = -\frac{k^2}{1+ \beta}\,,
\end{equation}
with $\beta = \frac{a_2}{a_1}$, as before. We now substitute
\begin{equation}
\tau = \frac{\frac{1}{a_1} + \frac{k^2}{a_1(1 + \beta)} \text{sn}^2 v}{1 - k^2 \text{sn}^2 v}\,,
\end{equation}
in (\ref{areaB^2<0}) and find
\begin{equation}
S = \frac{\pi l^2}{G_4}\frac{2 a_1}{\sqrt{1+2\beta}}\int \limits_{0}^{K(k)} dv\frac{\text{dn}^4 (v)}{(1-\alpha^2\text{sn}^2(v))^2}\ .
\label{ei.s}
\end{equation}
Apart from the coefficient and the limits, the indefinite integral
involved in (\ref{ei.s}) is the same as that in (\ref{ei.t}). The result
on substituting the integration limits is
\begin{equation}
  S = \frac{\pi l^2}{G_4}\frac{a_1}{\sqrt{1+2\beta}}
  \left[ -\beta K(k) + \left(1+2\beta\right)E(k)\right].
\label{s.int}
\end{equation}
This result only involves complete elliptic integrals. It is well defined
for the entire range of $\beta$, \ie\ $0 \le \beta \le 1$.

In the Nariai limit, we have $a_2 = a_1 = \frac{1}{\sqrt{3}}$\,, and\
$\beta \to 1 \,, \ k \to 0$,\ giving\ $K(0) = \frac{\pi}{2} =  E(0)$.\
This leads to the extremal limit of (\ref{s.int}),
\begin{equation}
S \to S_{\text{Nariai}} =  \frac{\pi^2 l^2}{3 G_4}\ .
\end{equation}

As mentioned after (\ref{areaB^2<0}), the de Sitter limit appears
slightly ill-defined. However one can formally calculate the area
(\ref{s.int}) above in this limit, with\ $\beta \to 0 \,, \ k \to
1$,\ and\ $K(1) \to \infty \,, \ E(1) = 1$,\ obtaining\ $S \to S_{dS}
= \frac{\pi l^2}{G_4}$\,, which is de Sitter entropy.



\begin{thebibliography}{} 

\footnotesize{

\bibitem{Ryu:2006bv} 
  S.~Ryu and T.~Takayanagi,
  ``Holographic derivation of entanglement entropy from AdS/CFT,''
  Phys.\ Rev.\ Lett.\  {\bf 96}, 181602 (2006)
  [hep-th/0603001].

\bibitem{Ryu:2006ef} 
  S.~Ryu and T.~Takayanagi,
  ``Aspects of Holographic Entanglement Entropy,''
  JHEP {\bf 0608}, 045 (2006)
  [hep-th/0605073].

\bibitem{HRT} 
V.~E.~Hubeny, M.~Rangamani and T.~Takayanagi,
``A Covariant holographic entanglement entropy proposal,'' 
JHEP {\bf 0707} (2007) 062  [arXiv:0705.0016 [hep-th]].

\bibitem{Rangamani:2016dms} 
  M.~Rangamani and T.~Takayanagi,
  ``Holographic Entanglement Entropy,''
  Lect.\ Notes Phys.\  {\bf 931}, pp.1 (2017)
  [arXiv:1609.01287 [hep-th]].

\bibitem{Maldacena:1997re}
  J.~M.~Maldacena,
  ``The large N limit of superconformal field theories and supergravity,''
  Adv.\ Theor.\ Math.\ Phys.\  {\bf 2}, 231 (1998)
  [Int.\ J.\ Theor.\ Phys.\  {\bf 38}, 1113 (1999)]
  [arXiv:hep-th/9711200].

\bibitem{Gubser:1998bc}
  S.~S.~Gubser, I.~R.~Klebanov and A.~M.~Polyakov,
  ``Gauge theory correlators from non-critical string theory,''
  Phys.\ Lett.\  B {\bf 428}, 105 (1998)
  [arXiv:hep-th/9802109].

\bibitem{Witten:1998qj}
  E.~Witten,
  ``Anti-de Sitter space and holography,''
  Adv.\ Theor.\ Math.\ Phys.\  {\bf 2}, 253 (1998)
  [arXiv:hep-th/9802150].

\bibitem{Aharony:1999ti}
  O.~Aharony, S.~S.~Gubser, J.~M.~Maldacena, H.~Ooguri and Y.~Oz,
  ``Large N field theories, string theory and gravity,''
  Phys.\ Rept.\  {\bf 323}, 183 (2000)
  [arXiv:hep-th/9905111].

\bibitem{Spradlin:2001pw} 
  M.~Spradlin, A.~Strominger and A.~Volovich,
  ``Les Houches lectures on de Sitter space,''
  hep-th/0110007.

\bibitem{Gibbons:1977mu} 
  G.~W.~Gibbons and S.~W.~Hawking,
  ``Cosmological Event Horizons, Thermodynamics, and Particle Creation,''
  Phys.\ Rev.\ D {\bf 15}, 2738 (1977).
  doi:10.1103/PhysRevD.15.2738

\bibitem{Strominger:2001pn} 
  A.~Strominger,
  ``The dS / CFT correspondence,''
  JHEP {\bf 0110}, 034 (2001)
  [hep-th/0106113].

\bibitem{Witten:2001kn} 
  E.~Witten,
  ``Quantum gravity in de Sitter space,''
  [hep-th/0106109].

\bibitem{Maldacena:2002vr}
  J.~M.~Maldacena,
  ``Non-Gaussian features of primordial fluctuations in single field inflationary models,''
  JHEP {\bf 0305}, 013 (2003),\ 
  [astro-ph/0210603].

\bibitem{Anninos:2011ui} 
  D.~Anninos, T.~Hartman and A.~Strominger,
  ``Higher Spin Realization of the dS/CFT Correspondence,''
  arXiv:1108.5735 [hep-th].

 \bibitem{Narayan:2017xca} 
 K.~Narayan,
 ``On extremal surfaces and de Sitter entropy,''
 Phys.\ Lett.\ B {\bf 779}, 214 (2018)
 [arXiv:1711.01107 [hep-th]].

\bibitem{Narayan:2019pjl} 
  K.~Narayan,
  ``de Sitter entropy as entanglement,''
  arXiv:1904.01223 [hep-th],\ \emph{Honorable Mention,\ Gravity Research
    Foundation 2019 Awards for Essays on Gravitation}, IJMPD 2019.

\bibitem{Narayan:2015vda} 
  K.~Narayan,
  ``de Sitter extremal surfaces,''
  Phys.\ Rev.\ D {\bf 91}, no. 12, 126011 (2015)
  [arXiv:1501.03019 [hep-th]]; \ \
  ``de Sitter space and extremal surfaces for spheres,''
  Phys.\ Lett.\ B {\bf 753}, 308 (2016)
  [arXiv:1504.07430 [hep-th]].\

\bibitem{Sato:2015tta} 
  Y.~Sato,
  ``Comments on Entanglement Entropy in the dS/CFT Correspondence,''
  Phys.\ Rev.\ D {\bf 91}, no. 8, 086009 (2015)
  [arXiv:1501.04903 [hep-th]].

\bibitem{Miyaji:2015yva} 
  M.~Miyaji and T.~Takayanagi,
  ``Surface/State Correspondence as a Generalized Holography,''
  PTEP {\bf 2015}, no. 7, 073B03 (2015)
  doi:10.1093/ptep/ptv089
  [arXiv:1503.03542 [hep-th]].

\bibitem{Hartman:2013qma} 
  T.~Hartman and J.~Maldacena,
  ``Time Evolution of Entanglement Entropy from Black Hole Interiors,''
  JHEP {\bf 1305}, 014 (2013)
  [arXiv:1303.1080 [hep-th]].

\bibitem{Maldacena:2001kr} 
  J.~M.~Maldacena,
  ``Eternal black holes in anti-de Sitter,''
  JHEP {\bf 0304}, 021 (2003)
  [hep-th/0106112].

\bibitem{Arias:2019pzy} 
  C.~Arias, F.~Diaz and P.~Sundell,
  ``De Sitter Space and Entanglement,''
  arXiv:1901.04554 [hep-th].

\bibitem{Narayan:2016xwq} 
  K.~Narayan,
  ``On $dS_4$ extremal surfaces and entanglement entropy in some ghost CFTs,''
  Phys.\ Rev.\ D {\bf 94}, no. 4, 046001 (2016)
  doi:10.1103/PhysRevD.94.046001
  [arXiv:1602.06505 [hep-th]]; \\ 
  D.~P.~Jatkar and K.~Narayan,
  ``Entangled spins and ghost-spins,''
  Nucl.\ Phys.\ B {\bf 922}, 319 (2017)
  [arXiv:1608.08351 [hep-th]]; \ \
  ``Ghost-spin chains, entanglement and $bc$-ghost CFTs,''
  Phys.\ Rev.\ D {\bf 96}, no. 10, 106015 (2017)
  [arXiv:1706.06828 [hep-th]];\\
  D.~P.~Jatkar, K.~S.~Kolekar and K.~Narayan,
  ``N-level ghost-spins and entanglement,''
  Phys.\ Rev.\ D {\bf 99}, 106003 (2019),
  arXiv:1812.07925 [hep-th].

\bibitem{Nariai}
  H. Nariai,
  ``On some static solutions of Einstein’s gravitational field equations
  in a spherically symmetric case'',
  Sci. Rep. Tohoku Univ. Eighth Ser. 34, 1950.

\bibitem{Anninos:2012ft} 
  D.~Anninos, F.~Denef and D.~Harlow,
  ``The Wave Function of Vasiliev's Universe - A Few Slices Thereof,''
  Phys.\ Rev.\ D {\bf 88}, 084049 (2013)
  [arXiv:1207.5517 [hep-th]].

\bibitem{Maldacena:2019cbz} 
  J.~Maldacena, G.~J.~Turiaci and Z.~Yang,
  ``Two dimensional Nearly de Sitter gravity,''
  arXiv:1904.01911 [hep-th].

\bibitem{Narayan:2020nsc}
K.~Narayan,
``de Sitter future-past extremal surfaces and the entanglement wedge,''
Phys. Rev. D \textbf{101}, no.8, 086014 (2020)
doi:10.1103/PhysRevD.101.086014
[arXiv:2002.11950 [hep-th]].

\bibitem{Ginsparg:1982rs} 
  P.~H.~Ginsparg and M.~J.~Perry,
  ``Semiclassical Perdurance of de Sitter Space,''
  Nucl.\ Phys.\ B {\bf 222}, 245 (1983).
  doi:10.1016/0550-3213(83)90636-3

\bibitem{Bousso:1995cc} 
  R.~Bousso and S.~W.~Hawking,
  ``The Probability for primordial black holes,''
  Phys.\ Rev.\ D {\bf 52}, 5659 (1995)
  doi:10.1103/PhysRevD.52.5659
  [gr-qc/9506047].

\bibitem{Bousso:1996au} 
  R.~Bousso and S.~W.~Hawking,
  ``Pair creation of black holes during inflation,''
  Phys.\ Rev.\ D {\bf 54}, 6312 (1996)
  doi:10.1103/PhysRevD.54.6312
  [gr-qc/9606052].

\bibitem{Das:2013mfa} 
  D.~Das, S.~R.~Das and K.~Narayan,
  ``dS/CFT at uniform energy density and a de Sitter 'bluewall',''
  JHEP {\bf 1404}, 116 (2014)
  doi:10.1007/JHEP04(2014)116
  [arXiv:1312.1625 [hep-th]].

\bibitem{Headrick:2010zt} 
  M.~Headrick,
  ``Entanglement Renyi entropies in holographic theories,''
  Phys.\ Rev.\ D {\bf 82}, 126010 (2010)
  [arXiv:1006.0047 [hep-th]].

\bibitem{Czech:2012bh} 
  B.~Czech, J.~L.~Karczmarek, F.~Nogueira and M.~Van Raamsdonk,
  ``The Gravity Dual of a Density Matrix,''
  Class.\ Quant.\ Grav.\  {\bf 29}, 155009 (2012)
  doi:10.1088/0264-9381/29/15/155009
  [arXiv:1204.1330 [hep-th]].

\bibitem{Wall:2012uf} 
  A.~C.~Wall,
  ``Maximin Surfaces, and the Strong Subadditivity of the Covariant Holographic Entanglement Entropy,''
  Class.\ Quant.\ Grav.\  {\bf 31}, no. 22, 225007 (2014)
  doi:10.1088/0264-9381/31/22/225007
  [arXiv:1211.3494 [hep-th]].

\bibitem{Headrick:2014cta} 
  M.~Headrick, V.~E.~Hubeny, A.~Lawrence and M.~Rangamani,
  ``Causality \& holographic entanglement entropy,''
  JHEP {\bf 1412}, 162 (2014)
  doi:10.1007/JHEP12(2014)162
  [arXiv:1408.6300 [hep-th]].

\bibitem{GuvenNunez}
  J.~Guven and D.~N\'u\~nez, ``Schwarzschild-de Sitter space and its perturbations,"
  Phys.\ Rev.\ D {\bf 42}, 2577 (1990)
  doi:10.1103/PhysRevD.42.2577.
  
\bibitem{Podolsky:1999ts} 
  J.~Podolsky,
  ``The Structure of the extreme Schwarzschild-de Sitter space-time,''
  Gen.\ Rel.\ Grav.\  {\bf 31}, 1703 (1999)
  doi:10.1023/A:1026762116655
  [gr-qc/9910029].

\bibitem{BF:1971}
P.~F.~Byrd and M.~D.~Friedmann, {\it Handbook of Elliptic Integrals for Engineers and Scientists}, Springer-Verlag Berlin Heidelberg (1971).
doi:10.1007/978-3-642-65138-0  
}




\end{thebibliography}
\end{document}